\documentclass[10pt]{article}
\usepackage{booktabs} 
\usepackage{caption}  
\usepackage{a4wide}
\usepackage{graphicx}
\usepackage{amssymb}
\usepackage{amsmath}
\usepackage{slashed}
\usepackage{comment}
\usepackage{cite}
\usepackage{color}
\usepackage{multicol}
\usepackage{braket}
\usepackage[colorlinks=true, linkcolor=red, citecolor=blue, urlcolor=blue]{hyperref}
\usepackage[affil-it]{authblk}
\usepackage[normalem]{ulem}
\usepackage{xcolor}
\newcommand{\be}{\begin{equation}}
\newcommand{\ee}{\end{equation}}
\newcommand{\ba}{\begin{eqnarray}}
\newcommand{\ea}{\end{eqnarray}}

\allowdisplaybreaks

\newcommand{\orcidicon}[1]{%
	\href{https://orcid.org/#1}{\includegraphics[height=8pt]{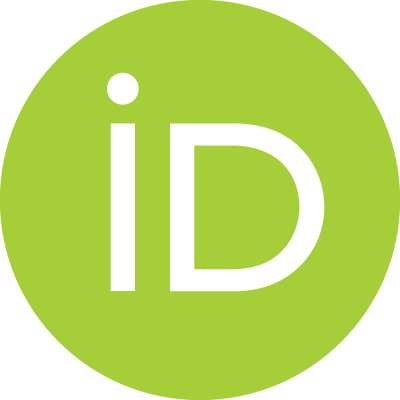}}%
}

\title{Semileptonic $\Omega_{b}^{*}\rightarrow\Omega_{c}^{*} \ell \bar{\nu}_{\ell}$ transition in QCD}

\author{\small
	A. Amiri$^1$\orcidicon{0000-0003-2479-4207}\thanks{\href{mailto:amir.amiri1308@gmail.com}{amir.amiri1308@gmail.com}},\ \
	P. Eslami$^1$\orcidicon{0000-0002-9308-2900}\thanks{\href{mailto:eslami@um.ac.ir}{eslami@um.ac.ir}; Corresponding author},\ \
	K. Azizi$^{2,3}$\orcidicon{0000-0003-3741-2167}\thanks{\href{mailto:kazem.azizi@ut.ac.ir}{kazem.azizi@ut.ac.ir}; Corresponding author},\ \
	R. Jafariseyedabad$^2$\orcidicon{0000-0002-9624-2295}\thanks{\href{mailto:jafari.reza13@gmail.com}{jafari.reza13@gmail.com}}}

\affil{\it $^1$Department of Physics, Faculty of Science, Ferdowsi University of Mashhad, P.O.Box 1436, Mashhad, Iran}
\vspace{-0.6cm}
\affil{\it $^2$Department of Physics, University of Tehran, North Karegar Avenue, Tehran 14395-547, Iran}
\vspace{-0.6cm}
\affil{\it $^3$Department of Physics, Dogus University, Dudullu-$\ddot{U}$mraniye, 34775 Istanbul, T$\ddot{u}$rkiye}

\date{}

\begin{document}

\maketitle

\begin{abstract}
We employ the QCD sum rule method to study the semileptonic weak decay of the single bottom baryon $\Omega_{b}^{*}$ with spin $\frac{3}{2}$ into the single charmed baryon $\Omega_{c}^{*}$ with spin $\frac{3}{2}$, corresponding to a $\frac{3}{2}\rightarrow\frac{3}{2}$ weak transition. A three-point correlation function is calculated in both the physical and theoretical sides to derive the sum rules for the form factors of the transition. The analysis incorporates both the perturbative and non-perturbative contributions up to mass dimension six. After determining the working regions of the auxiliary parameters and performing numerical calculations of the sum rules of the form factors, we extract the $q^2$-dependent fit functions for the form factors. The obtained fit functions are then applied to compute the decay widths of the $\Omega_{b}^{*}\rightarrow\Omega_{c}^{*} \ell \bar{\nu}_{\ell}$ transition in all lepton channels. Our results may serve as useful theoretical benchmarks for future experimental investigations of the semileptonic $\Omega_{b}^{*}\rightarrow\Omega_{c}^{*} \ell \bar{\nu}_{\ell}$ weak decays and the weak dynamics of excited heavy baryons.
  
\end{abstract}
\vfill

\vfill
{\footnotesize\noindent }


\vspace{7cm}

\tableofcontents

\section{Introduction}
Heavy hadron spectroscopy has recently become an important and active field due to the discovery of many new states. In this context, the study of single heavy baryons containing a $b$ or $c$ quark has attracted significant attention. Numerous theoretical studies have investigated the properties of heavy baryons, employing a wide range of models, methods, and decay modes \cite{Roberts:2007ni,Aliev:2008sk,Wang:2010vn,Wang:2009cr,Yoshida:2015tia,Agaev:2017jyt,Wang:2017kfr,Ortiz-Pacheco:2023kjn}. One important family of single heavy baryons is $\Omega_Q$, with $Q$ denoting either a $b$ or $c$ quark, which appear as spin-$\frac{1}{2}$ ($\Omega_Q$) or spin-$\frac{3}{2}$ ($\Omega^*_Q$) states. Different theoretical approaches, including QCD sum rules, lattice QCD and potential models have predicted the spectrum of the $\Omega_{c}^0$ baryons, including their narrow excited states \cite{Wang:2010vn,Wang:2009cr,Agaev:2017jyt,Wang:2017vnc,Padmanath:2017lng,Zhao:2017fov,Karliner:2017kfm,Patel:2025gbw}. On the other hand, in the past two decades experimental advancements have led to the observation of a wide range of new hadronic states, marking substantial progress in the development of hadron physics. The BaBar Collaboration has reported on the inclusive production and decays of the $\Omega_{c}^0$ baryons, where $\Omega_{c}^0$ have been reconstructed in the final states $\Omega^-\pi^+$, $\Omega^-\pi^+\pi^0$, $\Omega^-\pi^+\pi^+\pi^-$ and $\Xi^-K^-\pi^+\pi^+$ \cite{BaBar:2007jdg}. In addition, the Collaboration has reported the first observation of the excited singly-charmed baryon $\Omega_{c}^{*0}$, in the radiative decay $\Omega_{c}^0\gamma$, where the $\Omega_{c}^0$ decays into the aforementioned final states \cite{BaBar:2006pve}. Five new narrow excited states of the $\Omega_{c}^0$, $\Omega_{c}(3000)$, $\Omega_{c}(3050)$, $\Omega_{c}(3066)$, $\Omega_{c}(3090)$ and $\Omega_{c}(3119)$, have been observed by the LHCb \cite{LHCb:2017uwr} and the Belle Collaborations \cite{Belle:2017ext} in the $\varXi_c^+ K^-$ mass spectrum. They have also discussed possible interpretations of the spin and parity of these states \cite{LHCb:2021ptx}. Two additional $\Omega_{c}^0$ states, $\Omega_{c}(3185)$ and $\Omega_{c}(3327)$, were observed in the $\varXi_c^+ K^-$ invariant-mass spectrum using proton-proton collision data collected by the LHCb Collaboration \cite{LHCb:2023sxp}.
The quark model predicts fifteen heavy $b$-baryons, each consisting of a single $b$ quark and two lighter quarks ($u$, $d$, or $s$). Among them, four states, $\varLambda_b^0$, $\varXi_b^0$, $\varXi_b^-$ and $\Omega_{b}^-$, are of particular interest as they can decay only via the weak interaction. Heavy $b$-baryons have also been studied using different theoretical approaches, with investigations focusing on the spectra of both the ground and excited states through various models and methods \cite{Wang:2010vn,Wang:2009cr,Yoshida:2015tia,Agaev:2017jyt,Patel:2025gbw,BAGAN1992176,Ebert:2007nw,Wang:2009ozr,Vijande:2012mk,Chen:2016phw,Agaev:2017ywp,Liang:2017ejq,Xiao:2020oif}. Owing to the relatively large masses of $b$-baryons, only high-energy colliders, such as the LHC, can currently produce all their species in significant quantities. Among these four weakly decaying $b$-baryons, the heaviest $\Omega_{b}^-$, composed of $bss$ quarks, remains the least studied. The D$\emptyset$ and CDF Collaborations have independently reported the first observation of the doubly-strange baryon $\Omega_{b}^-$ via the decay channel $\Omega_{b}^-\rightarrow J/\psi\ \Omega^{-}$ in $p\bar{p}$ collisions \cite{D0:2008sbw,CDF:2009sbo}. In 2013, the LHCb Collaboration observed the $\Omega_{b}^-$ baryon using the same decay mode \cite{LHCb:2013wmn}, thereby confirming the CDF result \cite{CDF:2009sbo}. The CDF and LHCb experiments have also observed the $\Omega_{b}^-$ baryon through a second decay channel, $\Omega_{b}^-\rightarrow \Omega_c^0 \pi^-$ \cite{CDF:2014mon,LHCb:2016coe}. The first observation of the excited $\Omega_{b}^-$ states has been reported by the LHCb Collaboration. Four new narrow exited states $\Omega_{b}(6316)$, $\Omega_{b}(6330)$, $\Omega_{b}(6340)$ and $\Omega_{b}(6350)$, have been reconstructed in the near-threshold $\varXi_b^0 K^-$ resonances and their possible spin-parity assignments have also been discussed based on theoretical predictions \cite{LHCb:2020tqd}. Experimentally, the PDG \cite{ParticleDataGroup:2024cfk} reports the ground state $\Omega_{b}^-$ baryon, along with its four newly observed narrow excited states. 

The $\Omega_{b}^{*-}(1S,\frac{3}{2}^+)$ baryon is the anticipated hyperfine partner of the $\Omega_{b}^{-}(1S,\frac{1}{2}^+)$ state, with theoretical studies predicting a mass splitting of only about 10-40 MeV. From a theoretical standpoint, the $\Omega_{b}^*$ baryon occupies a distinctive position in the heavy-baryon spectrum. As a spin-$\frac{3}{2}$ ground state bottom baryon, it does not admit any Okubo-Zweig-Iizuka (OZI)-allowed strong decay channels. Furthermore, its radiative decay, $\Omega_{b}^{*}\rightarrow\Omega_{b}^{}\gamma$, involves the emission of a very soft photon, rendering its experimental observation particularly challenging~\cite{Peng:2024pyl}. These features make weak decay modes, and especially semileptonic transitions, the most promising probes of the internal structure and quantum numbers of the $\Omega_{b}^{*}$ baryon. In contrast to nonleptonic decays, semileptonic processes are theoretically cleaner, as they are free from final-state hadronic interactions and allow for direct access to the underlying weak transition form factors. Moreover, spin-$\tfrac{3}{2}\to\tfrac{3}{2}$ transitions possess a rich Lorentz structure characterized by a large set of independent form factors, which are tightly constrained in the heavy-quark limit by heavy-quark spin and flavor symmetries. Analyzing their structure, hierarchy, and relative magnitudes therefore provides a sensitive probe of heavy-quark symmetry relations and offers quantitative insight into symmetry-breaking effects arising from finite heavy-quark mass corrections. From the experimental perspective, semileptonic decays are particularly attractive due to their relatively clean signatures and reduced hadronic uncertainties compared to nonleptonic modes. With the steadily increasing statistics at LHCb and the prospects offered by future flavor-physics experiments, semileptonic decays of bottom baryons are becoming increasingly accessible. Altogether, these considerations underscore the relevance of the present analysis both for future experimental searches and for assessing the reliability of nonperturbative QCD approaches in heavy-baryon decays.

In this study, we investigate the semileptonic weak decay of the $\Omega_{b}^{*}$ baryon into the $\Omega_{c}^{*}$ baryon. In recent years, semileptonic decays have received considerable attention because of their potential to reveal deviations of the experimental data from the Standard Model (SM) predictions and to provide hints of possible new physics, which has become a central focus of both the theoretical and experimental studies, particularly following the discovery of the Higgs boson. Such decays also provide valuable insights into the internal structure of hadrons. This analysis is carried out using the QCD sum rule approach \cite{Shifman:1978bx,Shifman:1978by}, a reliable and powerful method that is widely employed in hadron spectroscopy and in the study of various decay modes \cite{Dehghan:2023ytx,Azizi:2024mmb,Najjar:2024deh,Neishabouri:2024gbc,Tousi:2024usi,Khajouei:2024frw}. By applying three-point QCD sum rules and considering the $b\rightarrow c$ transition at the quark level in the $\Omega_{b}^{*}\rightarrow\Omega_{c}^{*} \ell \bar{\nu}_{\ell}$ weak decay, we compute the corresponding form factors and the decay widths for all leptonic channels. 

The paper is organized as follows. In section \ref{QCDSumRules}, the QCD sum rule method and its basic concepts are described, including the evaluation of the three-point correlation function in two regions. Section \ref{NumCalc} presents the numerical analysis of the calculations, where the fitting functions of the form factors and the decay widths are obtained. The conclusions and a summary of the research are given in section \ref{con}. Finally, Appendix~\ref{ApenA} provides the final form of the correlation function in the theoretical side.  

\section{QCD Sum Rules}
\label{QCDSumRules}
The QCD sum rule approach is a powerful, non-perturbative method that allows the evaluation of the hadron-to-hadron transition matrix elements via a three-point correlation function at both the hadronic and quark-gluon levels. In the present study, we employ QCD sum rules to analyze the semileptonic weak decay of the $\Omega_{b}^{*}$ baryon and to investigate the properties of this transition. In the following, we discuss different aspects of the process.
\subsection{Correlation Function}
\label{Correlation Function}
In this study, the $\Omega_{b}^{*}$ baryon undergoes a semileptonic weak decay to $\Omega_{c}^{*} \ell \bar{\nu}_{\ell}$, proceeding via the $b\rightarrow c \ell \bar{\nu}_{\ell}$ transition. The $\Omega_{b}^{*}$ ($\Omega_{c}^{*}$) is a single heavy baryon containing one $b$ ($c$) heavy quark and two $ss$ light quarks. The properties of these baryons are summarized in Table~\ref{Baryons}.
\begin{table}[ht]
	\centering
	\caption{Quark content and quantum numbers of the baryons }
	\begin{tabular}{ccccc} 
		\toprule
		\textbf{baryon} & \textbf{quarks} & \textbf{spin-parity} & \textbf{charge} & \textbf{quark model} \\
		\midrule
		$\Omega_{b}^{*}$    & $(b s s)$    & $\frac{3}{2}^{+}$  & $-1$    & $sextet$             \\ 
		$\Omega_{c}^{*}$    & $(c s s)$    & $\frac{3}{2}^{+}$  & $0$    & $sextet$             \\ 
		\bottomrule
	\end{tabular}
	\label{Baryons}
\end{table}

To apply the QCD sum rule method to the semileptonic $\Omega_{b}^{*}\rightarrow\Omega_{c}^{*} \ell \bar{\nu}_{\ell}$ weak decay, we first evaluate a three-point correlation function, which is given by:
\begin{equation}
	\Pi_{\rho\mu\nu}(p,p',q^2)=i^2\int{d^4xe^{-ip.x}}\int{d^4ye^{ip'.y}\braket{0| \mathcal{T} \{\mathcal{J}_{\rho}^{\Omega_{c}^{*}}(y) \mathcal{J}_{\mu}^{tr}(0)\bar{\mathcal{J}}_{\nu}^{\Omega_{b}^{*}}(x)\} |0}}  \,, 
	\label{CorrF}
\end{equation}
where $\mathcal{T}$ represents the time-ordering operator and $\mathcal{J}_{\mu}^{tr}$, $\mathcal{J}_{\rho}^{\Omega_{b}^{*}}$ and $\mathcal{J}_{\nu}^{\Omega_{c}^{*}}$ are the transition current and the interpolating currents of the initial and final baryons, respectively. 

The three point correlation function, Eq.~(\ref{CorrF}), is a central element of the QCD sum rule approach. The hadron-to-hadron transition matrix element for the transition $\Omega_{b}^{*}\rightarrow\Omega_{c}^{*} \ell \bar{\nu}_{\ell}$ can be extracted from this correlation function. In the QCD sum rule method, the correlation function is evaluated in two distinct regions: the hadronic region, known as the physical (phenomenological) side, and the quark-gluon region, referred to as the theoretical (QCD) side. This distinction is reflected in the behavior of the three-point correlation function as a function of $q^2$. The variable $q^2$ can take both time-like (positive) and space-like (negative) values, corresponding to the physical and QCD sides, respectively. As $q^2$ changes from negative (QCD side) to positive (physical side) values, the correlation function begins to incorporate long distance quark-gluon interactions, during which quarks gradually combine to form hadrons.

To apply the QCD sum rule method in the calculation of Eq.~(\ref{CorrF}), several key considerations must be addressed. The transition current, $\mathcal{J}_{\mu}^{tr}$, consists of two components: the vector ($V^{\mu}$) and axial vector ($A^{\mu}$) currents. Since the weak interaction responsible for the $b \rightarrow c \ell \bar{\nu}_{\ell}$ transition is mediated by the $W$ boson, whose mass, $m_W\simeq80$ GeV, is significantly larger than the energy scale of such weak decays, the process can be effectively described by integrating out the $W$ boson. Consequently, this transition can be formulated through an effective Hamiltonian representing the corresponding quark-lepton interaction,
\begin{equation}
	\label{EffH}
	\mathcal{H}_{eff}  = \frac{G_F}{\sqrt{2}} \ V_{cb} \ [\bar{c}\ \gamma_{\mu}\ (1-\gamma_{5})\ b] \ [\bar{\ell}\ \gamma^{\mu}\ (1-\gamma_{5})\ \nu_{\ell}]   \,,
\end{equation}
where $V_{c b}$ originates from the CKM matrix in the SM and $G_F$ denotes the well-known Fermi coupling constant. The decay amplitude (matrix element) of the $\Omega_{b}^{*}\rightarrow\Omega_{c}^{*} \ell \bar{\nu}_{\ell}$ transition is obtained by inserting the effective Hamiltonian, Eq.~(\ref{EffH}), between the initial and final baryon states,
\begin{equation}
	\label{AmpT}
	\mathcal{M}  = \braket{\Omega_{c}^{*}|\ \mathcal{H}_{eff}\ |\Omega_{b}^{*}} =\frac{G_F}{\sqrt{2}} \ V_{cb} \ [\bar{\ell}\ \gamma^{\mu}\ (1-\gamma_{5})\ \nu_{\ell}] \braket{\Omega_{c}^{*}|\ [\bar{c}\ \gamma_{\mu}\ (1-\gamma_{5})\ b]\ |\Omega_{b}^{*}} \,.
\end{equation} 
As expected, the quark-hadron part of this amplitude contains both the vector and axial-vector contributions,
\begin{equation}
	\mathcal{M}^{quark-hadron}  \propto \braket{\Omega_{c}^{*}| \mathcal{J}_{\mu}^{tr,V-A}|\Omega_{b}^{*}}=\braket{\Omega_{c}^{*}|\ [\bar{c}\ \gamma_{\mu}\ (1-\gamma_{5})\ b]\ |\Omega_{b}^{*}} \,.
\end{equation} 

In the following, our aim is to evaluate Eq.~(\ref{CorrF}) in both the physical and QCD sides and to derive the QCD sum rules for the form factors of the $\Omega_{b}^{*}$ baryon weak decay.
\subsection{Phenomenological side}
In the physical side, hadrons are treated as complete entities in time-like region ($q^2 >0$). In this region, the hadron-to-hadron decay amplitude, Eq.~(\ref{AmpT}), for the $\Omega_{b}^{*}$ weak decay can be parameterized in terms of fourteen form factors, consistent with Lorentz invariance and parity requirements,

\begin{align} \label{AmTr}
		\mathcal{M}_{\mu}^V  =& \braket{\Omega_{c}^{*}|\ \mathcal{J}_{\mu}^{V} (\bar{c}\gamma_{\mu}b)\ |\Omega_{b}^{*}} = \bar{u}_{\Omega_{c}^{*}}^{\alpha}(p',s')\ [g_{\alpha\beta}(\gamma_{\mu}F_{1}(q^2)-i\sigma_{\mu\nu}\frac{q_\nu}{m_{\Omega_{b}^{*}}}F_{2}(q^2)+ \frac{q_\mu}{m_{\Omega_{b}^{*}}}F_{3}(q^2))\nonumber \\&+ \frac{q_{\alpha}q_{\beta}}{m_{\Omega_{b}^{*}}^2}\ (\gamma_{\mu}F_{4}(q^2)-i\sigma_{\mu\nu}\frac{q_\nu}{m_{\Omega_{b}^{*}}}F_{5}(q^2)+ \frac{q_\mu}{m_{\Omega_{b}^{*}}}F_{6}(q^2))\nonumber \\& + \frac{(g_{\alpha\mu}q_{\beta} - g_{\beta\mu} q_{\alpha})} {m_{\Omega_{b}^{*}}} F_{7}(q^2)]\ u_{\Omega_{b}^{*}}^{\beta}(p,s)  \,, \nonumber\\
		\mathcal{M}_{\mu}^A  =& \braket{\Omega_{c}^{*}|\ \mathcal{J}_{\mu}^{A} (\bar{c}\gamma_{\mu}\gamma_{5}b)\ |\Omega_{b}^{*}} = \bar{u}_{\Omega_{c}^{*}}^{\alpha}(p',s')\ [g_{\alpha\beta}(\gamma_{\mu}G_{1}(q^2)-i\sigma_{\mu\nu}\frac{q_\nu}{m_{\Omega_{b}^{*}}}G_{2}(q^2)+ \frac{q_\mu}{m_{\Omega_{b}^{*}}}G_{3}(q^2))\nonumber \\&+ \frac{q_{\alpha}q_{\beta}}{m_{\Omega_{b}^{*}}^2}\ (\gamma_{\mu}G_{4}(q^2)-i\sigma_{\mu\nu}\frac{q_\nu}{m_{\Omega_{b}^{*}}}G_{5}(q^2)+ \frac{q_\mu}{m_{\Omega_{b}^{*}}}G_{6}(q^2)) \nonumber \\&+ \frac{(g_{\alpha\mu}q_{\beta} - g_{\beta\mu} q_{\alpha})} {m_{\Omega_{b}^{*}}} G_{7}(q^2)]\ \gamma_5\ u_{\Omega_{b}^{*}}^{\beta}(p,s) \,,
\end{align} 
here seven form factors $F_{1}(q^2)$, $F_{2}(q^2)$, $F_{3}(q^2)$, $F_{4}(q^2)$, $F_{5}(q^2)$, $F_{6}(q^2)$, $F_{7}(q^2)$ describe the vector transition, while another seven form factors $G_{1}(q^2)$, $G_{2}(q^2)$, $G_{3}(q^2)$, $G_{4}(q^2)$, $G_{5}(q^2)$, $G_{6}(q^2)$, $G_{7}(q^2)$ correspond to the axial-vector transition \cite{Faessler:2009xn}. The variable $q$ is the transferred momentum to the leptons, defined as $q = p - p'$, with $p$ and $p'$ being the four-momenta of the initial and final baryons, respectively. $u_{\Omega_{b}^{*}}^{\alpha}(p,s)$ ($u_{\Omega_{c}^{*}}^{\alpha}(p',s')$) denotes the Rarita-Schwinger spinor of the initial (final) baryonic state with four momentum $p$ ($p'$) and spin $s$ ($s'$). For the initial and final spin-$\frac{3}{2}$ baryon states, the completeness (summation) relations over the corresponding Rarita-Schwinger spinors  are given by \cite{Aliev:2010ev},
\begin{equation}
	\label{RSSum}
	\begin{split}
		\sum_{s}u_{\beta}^{\Omega_{b}^{*}}(p,s)\ \bar{u}_{\nu}^{\Omega_{b}^{*}}(p,s)  =& -(\slashed{p}+ m_{\Omega_{b}^{*}})\ \{g_{\beta\nu}-\frac{1}{3}\gamma_{\beta}\gamma_{\nu}-\frac{2}{3}\frac{p_{\beta}p_{\nu}}{m_{\Omega_{b}^{*}}^{2}}+\frac{1}{3}\frac{p_{\beta}\gamma_{\nu}-p_{\nu}\gamma_{\beta}}{m_{\Omega_{b}^{*}}}\}  \,, \\
		\sum_{s'}u_{\rho}^{\Omega_{c}^{*}}(p',s')\ \bar{u}_{\alpha}^{\Omega_{c}^{*}}(p',s')  =& -(\slashed{p'}+ m_{\Omega_{c}^{*}})\ \{g_{\rho\alpha}-\frac{1}{3}\gamma_{\rho}\gamma_{\alpha}-\frac{2}{3}\frac{p'_{\rho}p'_{\alpha}}{m_{\Omega_{c}^{*}}^{2}}+\frac{1}{3}\frac{p'_{\rho}\gamma_{\alpha}-p'_{\alpha}\gamma_{\rho}}{m_{\Omega_{c}^{*}}}\} \,.
	\end{split}
\end{equation}

To derive QCD sum rules for the form factors in Eq.~(\ref{AmTr}), we first evaluate the three point correlation function, Eq.~(\ref{CorrF}), in the physical side. For this purpose, complete sets of intermediate hadronic states, carrying the same quantum numbers as the interpolating currents of the initial and final baryons, are inserted into Eq.~(\ref{CorrF}),
\begin{equation}
	\label{ComS}
	1=\ket{0}\bra{0}+\sum_{h}{\int{\frac{d^4p_h}{(2\pi)^4}(2\pi)\delta(p_{h}^2-m_{h}^2)\ket{h(p_h)}\bra{h(p_h)}}+ \mathrm{higher\ Fock\ states}} \,.
\end{equation} 

After inserting the complete sets of the initial and final baryonic states, Eq.~(\ref{ComS}), into the correlation function and carrying out the necessary algebraic manipulations, we obtain the following expression for the correlation function:
\begin{equation}
	\label{Corrgr}
	\Pi_{\rho\mu\nu}^{Phys.}(p,p',q^2)=\frac{\braket{0|\mathcal{J}_{\rho}^{\Omega_{c}^{*}}(0)|\Omega_{c}^{*}(p')}\braket{\Omega_{c}^{*}(p')|\mathcal{J}_{\mu}^{tr,V-A}(0)|\Omega_{b}^{*}(p)}\braket{\Omega_{b}^{*}(p)|\bar{\mathcal{J}}_{\nu}^{\Omega_{b}^{*}}(0)|0}}{(p^2-m_{\Omega_{b}^{*}}^{2})(p'^2-m_{\Omega_{c}^{*}}^{2})}+\ldots \,.
\end{equation}
It should be noted that the correlation function receives contributions from the ground state, excited states, and the continuum of the hadronic spectrum. The matrix element $\braket{0|\mathcal{J}^{\Omega_{x}^{*}}(0)|\Omega_{x}^{*}(p)}$ is referred to as the residue of the baryon, and is defined for the initial and final baryon states as follows,
\begin{equation}
	\label{Residue}
	\begin{aligned}[b]
	\braket{0|\mathcal{J}_{\rho}^{\Omega_{c}^{*}}(0)|\Omega_{c}^{*}(p')}=\lambda_{\Omega_{c}^{*}}\ u_{\rho}^{\Omega_{c}^{*}}(p',s') \,, \\
	\braket{\Omega_{b}^{*}(p)|\bar{\mathcal{J}}_{\nu}^{\Omega_{b}^{*}}(0)|0}=\lambda_{\Omega_{b}^{*}}^{\dagger}\ \bar{u}_{\nu}^{\Omega_{b}^{*}}(p,s) \,.
	\end{aligned}
\end{equation}

We substitute Eqs.~(\ref{AmTr}),~(\ref{RSSum}) and~(\ref{Residue}) into Eq.~(\ref{Corrgr}) to evaluate the correlation function in the physical side. This provides the initial representation of the physical side correlation function, expressed in terms of its Lorentz structures. At this step, several important points must be addressed. First, the Lorentz structures appearing in the correlation function on the physical side must be organized in a specific order, since not all of them are independent. In the present work, the Dirac matrices are arranged in the order $\gamma_{\rho}\gamma_{\mu}\slashed{p}\slashed{p'}\gamma_{\nu}\gamma_{5}$, which allows dependent Lorentz structures to merge into an independent tensor basis. Second, since we are studying a $\frac{3}{2}\rightarrow\frac{3}{2}$ transition, the interpolating currents of the initial and final baryons can also couple to the spin-$\frac{1}{2}$ states in addition to the desired spin-$\frac{3}{2}$ states. These unwanted spin-$\frac{1}{2}$ contributions must be removed. In general, such contributions can be expressed as \cite{Aliev:2008sk,Aliev:2010ev}, 
\begin{equation}
		\braket{0|\mathcal{J}_{\mu}|p,s=\frac{1}{2}}=(A\gamma_{\mu}+Bp_{\mu})\ u(p,s=\frac{1}{2}) \,.
\end{equation}
By imposing the conditions $\mathcal{J}_{\rho}\gamma^{\rho}=0$ and $\mathcal{J}_{\nu}\gamma^{\nu}=0$ for the initial and final states, the spin-$\frac{1}{2}$ contributions can be defined as follow,
\begin{equation}
	\begin{split}
		\braket{0|\mathcal{J}_{\nu}^{\Omega_{b}^{*}}|p,s = \frac{1}{2}} =&\ (A\gamma_{\nu}-\frac{4A}{m}\ p_{\nu})\ u(p,s=\frac{1}{2}) \,, \\
		\braket{0|\mathcal{J}_{\rho}^{\Omega_{c}^{*}}|p',s'= \frac{1}{2}} =&\ (A'\gamma_{\rho}-\frac{4A'}{m'}\ p'_{\rho})\ u(p',s'=\frac{1}{2}) \,.
	\end{split}
\end{equation}
To isolate the desired spin-$\tfrac{3}{2}\to\tfrac{3}{2}$ contributions and suppress the unwanted spin-$\tfrac{1}{2}$ contaminations, one must remove the structures containing $\gamma_{\rho}$ at the far left and $\gamma_{\nu}$ or $\gamma_{\nu}\gamma_5$ at the far right, as well as the terms proportional to $p'_{\rho}$ and $p_{\nu}$ after applying the above ordering prescription. The remaining Lorentz structures are therefore free from the explicit spin-$\tfrac{1}{2}$ contaminations and are associated with the desired spin-$\tfrac{3}{2}\to\tfrac{3}{2}$ transition. Third, as mentioned earlier, the correlation function also receives contributions from the excited states and the hadronic continuum. To suppress these contributions, a double Borel transformation is applied to the correlation function of the physical side \cite{Aliev:2006gk},
\begin{equation}
\label{BorelT}
\hat{B} \frac{1}{(p^2-m^2)^n}\frac{1}{(p'^2-m'^2)^m}\rightarrow (-1)^{n+m} \frac{1}{\Gamma[n]\Gamma[m]} \frac{1}{(M^2)^{n-1}} \frac{1}{(M'^2)^{m-1}} e^{-\frac{m^2}{M^2}} e^{-\frac{m'^2}{M'^2}} \,,
\end{equation}
where $M^2$ and $M'^2$ denote the Borel mass parameters, while $m$ and $m'$ correspond to the masses of the initial and final baryons, respectively.

By implementing the three aforementioned steps, the final form of the correlation function in the physical side is determined as,  
\begin{align}
		&\Pi_{\rho\mu\nu}^{Phys.}(p,p',q^2)=\lambda_{\Omega_{b}^{*}} \lambda_{\Omega_{c}^{*}} e^{-\frac{m_{\Omega_{b}^{*}}^{2}}{M^2}} e^{-\frac{m_{\Omega_{c}^{*}}^{2}}{M'^2}}[F_1(q^2)(m_{\Omega_{c}^*}g_{\nu\rho}\gamma_{\mu}\slashed{p}- m_{\Omega_{b}^*}g_{\nu\rho}\gamma_{\mu}\slashed{p'}+g_{\nu\rho}\gamma_{\mu}\slashed{p} \slashed{p'}) \nonumber \\& +F_2(q^2)((m_{\Omega_{c}^*} +\frac{m_{\Omega_{c}^*}^{2}}{m_{\Omega_{b}^*}}) g_{\nu\rho} \gamma_{\mu} \slashed{p} - (m_{\Omega_{b}^*}+m_{\Omega_{c}^*})g_{\nu\rho}\gamma_{\mu}\slashed{p'} + (1+\frac{m_{\Omega_{c}^*}}{m_{\Omega_{b}^*}}) g_{\nu\rho}\gamma_{\mu}\slashed{p} \slashed{p'} - p_\mu g_{\nu\rho} \slashed{p'}-p'_\mu g_{\nu\rho} \slashed{p'}+\frac{1}{m_{\Omega_{b}^*}}p_\mu g_{\nu\rho} \slashed{p}\slashed{p'} \nonumber \\&+\frac{1}{m_{\Omega_{b}^*}}p'_\mu g_{\nu\rho} \slashed{p}\slashed{p'}) +F_3(q^2)(p_\mu g_{\nu\rho} \slashed{p'}-p'_\mu g_{\nu\rho} \slashed{p'}-\frac{1}{m_{\Omega_{b}^*}}p_\mu g_{\nu\rho} \slashed{p}\slashed{p'}+ \frac{1}{m_{\Omega_{b}^*}}p'_\mu g_{\nu\rho} \slashed{p}\slashed{p'})  +F_4(q^2)(-\frac{m_{\Omega_{c}^*}}{m_{\Omega_{b}^*}^2}p_\rho p'_\nu \gamma_{\mu} \slashed{p} \nonumber \\&+ \frac{1}{m_{\Omega_{b}^*}}p_\rho p'_\nu \gamma_{\mu} \slashed{p'} - \frac{1}{m_{\Omega_{b}^*}^2}p_\rho p'_\nu \gamma_{\mu} \slashed{p} \slashed{p'}) + F_5(q^2)(-(\frac{m_{\Omega_{c}^*}}{m_{\Omega_{b}^*}^2}+\frac{m_{\Omega_{c}^*}^2}{m_{\Omega_{b}^*}^3})p_\rho p'_\nu \gamma_{\mu} \slashed{p} + (\frac{1}{m_{\Omega_{b}^*}} + \frac{m_{\Omega_{c}^*}}{m_{\Omega_{b}^*}^2})p_\rho p'_\nu \gamma_{\mu} \slashed{p'} \nonumber \\&- (\frac{1}{m_{\Omega_{b}^*}^2}+\frac{m_{\Omega_{c}^*}}{m_{\Omega_{b}^*}^3})p_\rho p'_\nu \gamma_{\mu} \slashed{p} \slashed{p'}+ \frac{m_{\Omega_{c}^*}}{m_{\Omega_{b}^*}^3} p_\mu p_\rho p'_\nu \slashed{p}+ \frac{1}{m_{\Omega_{b}^*}^2}p_\mu p_\rho p'_\nu \slashed{p'}+\frac{1}{m_{\Omega_{b}^*}^2} p_\rho p'_\mu p'_\nu \slashed{p'}) +F_6(q^2)(-\frac{m_{\Omega_{c}^*}}{m_{\Omega_{b}^*}^3} p_\mu p_\rho p'_\nu \slashed{p} \nonumber \\&-\frac{1}{m_{\Omega_{b}^*}^2}p_\mu p_\rho p'_\nu \slashed{p'} +\frac{1}{m_{\Omega_{b}^*}^2} p_\rho p'_\mu p'_\nu \slashed{p'}) +F_7(q^2)(-p_\rho g_{\mu\nu} \slashed{p'}-p'_\nu g_{\mu\rho}\slashed{p'}+ \frac{1}{m_{\Omega_{b}^*}} p_\rho g_{\mu\nu} \slashed{p} \slashed{p'} + \frac{1}{m_{\Omega_{b}^*}} p'_\nu g_{\mu\rho}\slashed{p}\slashed{p'}) \nonumber \\& + G_1(q^2)(m_{\Omega_{c}^*}g_{\nu\rho}\gamma_{\mu}\slashed{p}\gamma_{5}+ m_{\Omega_{b}^*}g_{\nu\rho}\gamma_{\mu}\slashed{p'}\gamma_{5}+g_{\nu\rho}\gamma_{\mu}\slashed{p} \slashed{p'}\gamma_{5}) +G_2(q^2)((\frac{m_{\Omega_{c}^*}^{2}}{m_{\Omega_{b}^*}}-m_{\Omega_{c}^*}) g_{\nu\rho} \gamma_{\mu} \slashed{p}\gamma_{5} + (m_{\Omega_{c}^*}-m_{\Omega_{b}^*})g_{\nu\rho}\gamma_{\mu}\slashed{p'}\gamma_{5} \nonumber \\& - (1-\frac{m_{\Omega_{c}^*}}{m_{\Omega_{b}^*}}) g_{\nu\rho}\gamma_{\mu}\slashed{p} \slashed{p'}\gamma_{5} + p_\mu g_{\nu\rho} \slashed{p'}\gamma_{5}+p'_\mu g_{\nu\rho} \slashed{p'}\gamma_{5} +\frac{1}{m_{\Omega_{b}^*}}p_\mu g_{\nu\rho} \slashed{p}\slashed{p'}\gamma_{5}  +\frac{1}{m_{\Omega_{b}^*}}p'_\mu g_{\nu\rho} \slashed{p}\slashed{p'}\gamma_{5}) +G_3(q^2)(-p_\mu g_{\nu\rho} \slashed{p'}\gamma_{5} \nonumber \\& +p'_\mu g_{\nu\rho} \slashed{p'}\gamma_{5} -\frac{1}{m_{\Omega_{b}^*}}p_\mu g_{\nu\rho} \slashed{p}\slashed{p'}\gamma_{5}+ \frac{1}{m_{\Omega_{b}^*}}p'_\mu g_{\nu\rho} \slashed{p}\slashed{p'}\gamma_{5})  +G_4(q^2)(-\frac{m_{\Omega_{c}^*}}{m_{\Omega_{b}^*}^2}p_\rho p'_\nu \gamma_{\mu} \slashed{p}\gamma_{5} - \frac{1}{m_{\Omega_{b}^*}}p_\rho p'_\nu \gamma_{\mu} \slashed{p'}\gamma_{5} \nonumber \\& - \frac{1}{m_{\Omega_{b}^*}^2}p_\rho p'_\nu \gamma_{\mu} \slashed{p} \slashed{p'}\gamma_{5}) + G_5(q^2)((\frac{m_{\Omega_{c}^*}}{m_{\Omega_{b}^*}^2}-\frac{m_{\Omega_{c}^*}^2}{m_{\Omega_{b}^*}^3})p_\rho p'_\nu \gamma_{\mu} \slashed{p}\gamma_{5} + (\frac{1}{m_{\Omega_{b}^*}} - \frac{m_{\Omega_{c}^*}}{m_{\Omega_{b}^*}^2})p_\rho p'_\nu \gamma_{\mu} \slashed{p'}\gamma_{5} + (\frac{1}{m_{\Omega_{b}^*}^2}-\frac{m_{\Omega_{c}^*}}{m_{\Omega_{b}^*}^3})p_\rho p'_\nu \gamma_{\mu} \slashed{p} \slashed{p'}\gamma_{5} \nonumber \\& + \frac{m_{\Omega_{c}^*}}{m_{\Omega_{b}^*}^3} p_\mu p_\rho p'_\nu \slashed{p}\gamma_{5} - \frac{1}{m_{\Omega_{b}^*}^2}p_\mu p_\rho p'_\nu \slashed{p'}\gamma_{5} -\frac{1}{m_{\Omega_{b}^*}^2} p_\rho p'_\mu p'_\nu \slashed{p'}\gamma_{5}) +G_6(q^2)(-\frac{m_{\Omega_{c}^*}}{m_{\Omega_{b}^*}^3} p_\mu p_\rho p'_\nu \slashed{p}\gamma_{5}  +\frac{1}{m_{\Omega_{b}^*}^2}p_\mu p_\rho p'_\nu \slashed{p'}\gamma_{5} \nonumber \\& -\frac{1}{m_{\Omega_{b}^*}^2} p_\rho p'_\mu p'_\nu \slashed{p'}\gamma_{5}) +G_7(q^2)(p_\rho g_{\mu\nu} \slashed{p'}\gamma_{5} +p'_\nu g_{\mu\rho}\slashed{p'}\gamma_{5}+ \frac{1}{m_{\Omega_{b}^*}} p_\rho g_{\mu\nu} \slashed{p} \slashed{p'} \gamma_{5} + \frac{1}{m_{\Omega_{b}^*}} p'_\nu g_{\mu\rho}\slashed{p} \slashed{p'} \gamma_{5})]+\ldots  \,. 
\end{align}
\subsection{Theoretical side}
In the QCD side, the constituents of hadrons and their interactions must be taken into account. Hadrons are described in terms of quarks and gluons, with their interactions in space-like region ($q^2 < 0$). In this region, the correlation function can be computed by applying the Operator Product Expansion (OPE) in the deep Euclidean domain, where $q^2\ll-\Lambda_{QCD}^{2}$. To evaluate the correlation function, Eq.~(\ref{CorrF}), in the QCD side, both the perturbative and non-perturbative contributions must be taken into account. In the following, we outline how the correlation function of the QCD side can be computed in terms of the light and heavy quark propagators, and how these propagators are employed to systematically calculate the perturbative and non-perturbative contributions.

To evaluate the correlation function, one must employ the interpolating currents corresponding to the initial and final baryonic states. Since both baryons are single heavy states with spin-$\frac{3}{2}$, their interpolating currents are given by \cite{Aliev:2008sk},
\begin{equation}
\label{IntPol}
\mathcal{J}_{\mu}^{\Omega_{Q}^{*}}(x)=\sqrt{\frac{1}{3}}\epsilon_{abc}\{(s^{aT}(x)C\gamma_{\mu}s^{b}(x))Q^{c}(x)+(s^{aT}(x)C\gamma_{\mu}Q^b(x))s^{c}(x)+(Q^{aT}(x)C\gamma_{\mu}s^{b}(x))s^{c}(x)\} \,,
\end{equation}
where $a$, $b$, and $c$ are color indices, $C$ denotes the charge conjugation operator, and $s(x)$ and $Q(x)$ represent the \textit{strange} and heavy quark fields, respectively. 

We substitute the interpolating currents of the initial and final baryons, Eq.~(\ref{IntPol}), into Eq.~(\ref{CorrF}) and calculate the correlation function in the theoretical side. By contracting the relevant quark fields, the correlation function is acquired in terms of the light and heavy quark propagators,
\begin{align}
\label{QCDCorr}
&\Pi_{\rho\mu\nu}^{QCD}(p,p',q^2)= i^2\int{d^4x\ e^{-ip.x}}\int{d^4y\ e^{ip'.y}}\ \frac{1}{3}\ \epsilon_{abc}\ \epsilon_{a'b'c'}\ \{\nonumber \\& Tr[\gamma_{\rho}S_s^{b'b}(y-x)\gamma_{\nu}\tilde{S}_s^{a'a}(y-x)]S_c^{c'i}(y)\gamma_{\mu}(1-\gamma_{5}) S_b^{ic}(-x) - Tr[\gamma_{\rho}S_s^{b'a}(y-x)\gamma_{\nu}\tilde{S}_s^{a'b}(y-x)]S_c^{c'i}(y)\gamma_{\mu}(1-\gamma_{5}) S_b^{ic}(-x) \nonumber \\& - S_c^{c'i}(y)\gamma_{\mu}(1-\gamma_{5}) S_b^{ib}(-x)\gamma_{\nu}\tilde{S}_s^{a'a}(y-x) \gamma_{\rho}S_s^{b'c}(y-x) + S_c^{c'i}(y)\gamma_{\mu}(1-\gamma_{5}) S_b^{ib}(-x)\gamma_{\nu} \tilde{S}_s^{b'a}(y-x)\gamma_{\rho}S_s^{a'c}(y-x) \nonumber \\& - S_c^{c'i}(y)\gamma_{\mu}(1-\gamma_{5}) S_b^{ia}(-x)\gamma_{\nu}\tilde{S}_s^{b'b}(y-x) \gamma_{\rho}S_s^{a'c}(y-x) + S_c^{c'i}(y)\gamma_{\mu}(1-\gamma_{5}) S_b^{ia}(-x)\gamma_{\nu} \tilde{S}_s^{a'b}(y-x)\gamma_{\rho} S_s^{b'c}(y-x) \nonumber \\& + S_s^{c'a}(y-x) \gamma_{\nu}\tilde{S}_s^{a'b}(y-x) \gamma_{\rho} S_c^{b'i}(y) \gamma_{\mu}(1-\gamma_{5}) S_b^{ic}(-x) - S_s^{c'b}(y-x) \gamma_{\nu}\tilde{S}_s^{a'a}(y-x) \gamma_{\rho} S_c^{b'i}(y) \gamma_{\mu}(1-\gamma_{5})S_b^{ic}(-x) \nonumber \\& - S_s^{c'a}(y-x) \gamma_{\nu}\tilde{S}_b^{ib}(-x) (1-\gamma_{5})\gamma_{\mu}\tilde{S}_c^{b'i}(y)\gamma_{\rho} S_s^{a'c}(y-x) + Tr[\gamma_{\rho} S_c^{b'i}(y) \gamma_{\mu}(1-\gamma_{5}) S_b^{ib}(-x)\gamma_{\nu}\tilde{S}_s^{a'a}(y-x)] S_s^{c'c}(y-x) \nonumber \\& + S_s^{c'b}(y-x) \gamma_{\nu}\tilde{S}_b^{ia}(-x) (1-\gamma_{5})\gamma_{\mu}\tilde{S}_c^{b'i}(y)\gamma_{\rho} S_s^{a'c}(y-x) - Tr[\gamma_{\rho} S_c^{b'i}(y) \gamma_{\mu}(1-\gamma_{5}) S_b^{ia}(-x)\gamma_{\nu} \tilde{S}_s^{a'b}(y-x)] S_s^{c'c}(y-x) \nonumber \\& - S_s^{c'a}(y-x) \gamma_{\nu}\tilde{S}_s^{b'b}(y-x) \gamma_{\rho} S_c^{a'i}(y) \gamma_{\mu}(1-\gamma_{5})S_b^{ic}(-x) + S_s^{c'b}(y-x) \gamma_{\nu}\tilde{S}_s^{b'a}(y-x) \gamma_{\rho} S_c^{a'i}(y) \gamma_{\mu}(1-\gamma_{5})S_b^{ic}(-x) \nonumber \\& + S_s^{c'a}(y-x) \gamma_{\nu} \tilde{S}_b^{ib}(-x) (1-\gamma_{5})\gamma_{\mu}\tilde{S}_c^{a'i}(y)\gamma_{\rho} S_s^{b'c}(y-x) - Tr[\gamma_{\mu}(1-\gamma_{5})S_b^{ib}(-x) \gamma_{\nu} \tilde{S}_s^{b'a}(y-x) \gamma_{\rho} S_c^{a'i}(y)] S_s^{c'c}(y-x) \nonumber \\& - S_s^{c'b}(y-x) \gamma_{\nu}\tilde{S}_b^{ia}(-x) (1-\gamma_{5})\gamma_{\mu}\tilde{S}_c^{a'i}(y)\gamma_{\rho} S_s^{b'c}(y-x) + Tr[\gamma_{\mu}(1-\gamma_{5})S_b^{ia}(-x) \gamma_{\nu} \tilde{S}_s^{b'b}(y-x) \gamma_{\rho} S_c^{a'i}(y)] S_s^{c'c}(y-x) \} \,,
\end{align}
where $\tilde{S}_{q}(x)=CS_q^T(x)C$. We compute both the perturbative and non-perturbative contributions to the QCD side of the correlation function by employing these propagators. The light and heavy quark propagators are represented as follows \cite{Agaev:2020zad}.  
\begin{equation}
\label{LHQPro}
\begin{split}
&S_{q}^{ab}(x) = i\delta_{ab}\frac{\slashed{x}} {2\pi^{2}x^{4}} -\delta_{ab}\frac{m_{q}}{4\pi^{2}x^{2}}-\delta_{ab}\frac{\langle \bar{q}q \rangle}{12}+i\delta_{ab}\frac{\slashed{x}m_{q}\langle \bar{q}q \rangle}{48}-\delta_{ab}\frac{x^2}{192}\langle \bar{q}g_{s}\sigma G q \rangle +i\delta_{ab}\frac{x^2\slashed{x}m_{q}}{1152}\langle \bar{q}g_{s}\sigma G q \rangle \\&-i\frac{g_{s}G_{ab}^{\mu\nu}}{32\pi^{2}x^{2}}[\slashed{x}\sigma_{\mu\nu} +\sigma_{\mu\nu}\slashed{x}]-i\delta_{ab}\frac{x^2\slashed{x}g_{s}^{2}\langle \bar{q}q \rangle^{2}}{7776}+ \ldots  \,, \\
&S_{Q}^{ab}(x) = i\int \frac{d^4k}{(2\pi)^{4}} e^{-ikx} \{\frac{\delta_{ab}(\slashed{k}+m_{Q})}{k^2-m_{Q}^{2}} - \frac{g_{s}G_{ab}^{\mu\nu}}{4} \frac{\sigma_{\mu\nu}(\slashed{k}+m_{Q})+(\slashed{k}+m_{Q})\sigma_{\mu\nu}}{(k^2-m_{Q}^{2})^2}\\&   +\delta_{ab}\ m_{Q} \frac{g_{s}^{2} G^{2}}{12} \frac{k^2+m_{Q} \slashed{k}}{(k^2-m_{Q}^{2})^4}+ \delta_{ab} \frac{g_{s}^{3} G^{3}}{48} \frac{(\slashed{k}+ m_{Q})}{(k^2-m_{Q}^{2})^{6}}[\slashed{k}(k^{2}-3 m_{Q}^{2})+2 m_{Q} (2 k^{2}-m_{Q}^{2})](\slashed{k}+m_{Q})+ \ldots \} \,, 
\end{split}
\end{equation}
where $k$ denotes the four momentum of heavy quark. The light and heavy quark propagators incorporate both the perturbative and non-perturbative effects. The perturbative contributions correspond to the terms with mass dimension $d = 0$ in both propagators. The non-perturbative contributions, commonly referred to as \textit{vacuum condensates}, involve operators of mass dimensions three ($d=3$), four ($d=4$), five ($d=5$), and six ($d=6$). These operators include the quark condensate, $\langle \bar{q}q \rangle$, gluon condensate, $\langle G^2 \rangle$, quark-gluon condensate, $\langle \bar{q}g_{s}\sigma G q \rangle$, four quark condensate, $\langle \bar{q}q \rangle^2$, and three gluon condensate, $\langle G^3 \rangle$. 

To extract the different structures of the QCD side of the correlation function, both the perturbative and non-perturbative contributions must be evaluated. These contributions are obtained by substituting the light and heavy quark propagators, Eq.~(\ref{LHQPro}), into the correlation function, Eq.~(\ref{QCDCorr}), which generates various terms corresponding to the perturbative and non-perturbative effects. In the present work, the correlation function of the QCD side is calculated by including the non-perturbative contributions up to mass dimension six, with all possible permutations taken into account.

The following relations are also employed in the calculations \cite{Najjar:2024deh},
\begin{equation}
G_{ab}^{\mu\nu}=G_{A}^{\mu\nu} \lambda_{ab} ^{A}/2,\ G^2=G_{\mu\nu}^{A}G_{A}^{\mu\nu},\ G^3 = f^{ABC} G_{A}^{\mu\nu}G_{B}^{\nu\lambda}G_{C}^{\lambda\mu} \,, 
\end{equation}
\begin{equation}
\braket{0| G_{\mu\nu}^{A}G_{\mu'\nu'}^{B} |0}= \frac{\langle G^2 \rangle}{96} \delta^{AB} [g_{\mu\mu'}g_{\nu\nu'}-g_{\mu\nu'}g_{\mu'\nu}] \,,
\end{equation}
\begin{equation}
t^{ab}t^{a'b'}= \frac{1}{2} (\delta^{ab'} \delta^{a'b}- \frac{1}{3}\delta^{ab} \delta^{a'b'}) \,,
\end{equation}
\begin{equation}
\begin{aligned}[b]
\braket{0| G_{\mu\nu}^{A} G_{\mu'\nu'}^{B} G_{\mu''\nu''}^{C} |0} &= \frac{\langle G^3 \rangle}{576} f^{ABC} [g_{\mu'\nu}g_{\mu''\nu'}g_{\nu''\mu}-g_{\nu''\mu}g_{\nu\nu'}g_{\mu'\mu''}-g_{\mu\mu''}g_{\mu'\nu}g_{\nu'\nu''}+g_{\mu\mu''}g_{\nu\nu'}g_{\mu'\nu''} \\& +g_{\mu\mu'}g_{\nu\mu''}g_{\nu'\nu''}-g_{\mu\mu'}g_{\nu\nu''}g_{\mu''\nu'}-g_{\mu\nu'}g_{\nu\mu''}g_{\nu''\mu'}+g_{\mu\nu'}g_{\nu\nu''}g_{\mu'\mu''}] \,,
\end{aligned}
\end{equation}
\begin{equation}
f^{ABC}t_{A}^{ab}t_{B}^{a'b'}t_{C}^{a''b''}= \frac{18}{56} i [\delta^{ba'} \delta^{b'a''}\delta^{b''a}- \frac{1}{3}(\delta^{ba'} \delta^{ab'}\delta^{a''b''}+\delta^{a'b''} \delta^{b'a''}\delta^{ab}+\delta^{ab''} \delta^{ba''}\delta^{a'b'})+\frac{2}{9}\delta^{ab} \delta^{a'b'}\delta^{a''b''}] \,,
\end{equation}
here $f^{ABC}$ are the structure constants of the $SU_{c}(3)$ gauge group of the QCD Lagrangian and $t^A = \lambda^A/2$ with $\lambda^A$, being the Gell-Mann matrices, where A, B, C = 1, 2 ... 8. The gluon field strength tensor is evaluated at the origin, i.e. $G_{\mu\nu}^A=G_{\mu\nu}^A(0)$.

After substituting the light and heavy quark propagators, including both the perturbative and non-perturbative terms, into Eq.~(\ref{QCDCorr}) in coordinate space, the resulting expressions contain four integrals and Fourier integrals for each term. These integrals, which arise with different integrands in the perturbative and non-perturbative parts of the correlation function, must be evaluated carefully. For this purpose, we employed a set of standard identities, formulas, and relations. Before proceeding with further calculations, a few remarks are in order. During the evaluation of the perturbative and non-perturbative contributions, most terms acquire imaginary parts. In the present study, we note that only the non-perturbative contributions of mass dimension six do not contain imaginary parts, whereas the lower dimensional terms do.

To extract the imaginary contributions of the relevant terms, the following procedure is employed. Using the given identity in Eq.~(\ref{FiIden}), parts of the denominators are expressed in exponential form \cite{Azizi:2017ubq},
\begin{equation}
\label{FiIden}
\frac{1}{[(y-x)^2]^n}=i \int \frac{d^{D}t}{(2\pi)^2} e^{-it.(y-x)} (-1)^{n+1} 2^{D-2n} \pi^{D/2} \frac{\Gamma(D/2-n)}{\Gamma(n)}(-\frac{1}{t^2})^{D/2-n} \,.
\end{equation}
The Fourier integrals are evaluated using the relation,
\begin{equation}
\int d^{4}x e^{i(k-p+t).x} \int d^{4}y e^{i(-k'+p'-t).y} = (2\pi)^4 \delta^4(k-p+t) (2\pi)^4 \delta^4(-k'+p'-t) \,.
\end{equation}
The Feynman parametrization is also applied to solve the four integrals, which allows us to combine multiple propagator denominators into a single denominator. For three propagators the parametrization is given by \cite{Najjar:2024deh},
\begin{equation}
\frac{1}{A^{n}B^{n'}C^{n''}}= \frac{\Gamma(n+n'+n'')}{\Gamma(n)\Gamma(n')\Gamma(n'')} \int_{0}^{1} \int_{0}^{1-r} dz dr \frac{r^{n-1}z^{n'-1}(1-r-z)^{n''-1}}{[rA+zB+(1-r-z)C]^{n+n'+n''}} \,.
\end{equation}
The final integrals over the parameter $t$ are evaluated by applying the relation \cite{Azizi:2017ubq},
\begin{equation}
\int d^{D}t \frac{(t^2)^n}{(t^2+\Delta)^m}  = \frac{i\pi^2(-1)^{n-m}\Gamma(n+2)\Gamma(m-n-2)}{\Gamma(2)\Gamma(m)[-\Delta]^{m-n-2}} \,.
\end{equation}
After evaluating the integrals, the imaginary parts of the obtained results are extracted by employing the identity \cite{Azizi:2017ubq},
\begin{equation}
\Gamma[\frac{D}{2}-n](-\frac{1}{\Delta})^{D/2-n} = \frac{(-1)^{n-1}}{(n-2)!} (-\Delta)^{n-2}\  ln(-\Delta) \,.
\end{equation}
These imaginary parts contribute to the \textit{spectral density}, defined as $\rho_i(s)=\frac{1}{\pi} Im[\Pi_{i}^{QCD}]$. In the present study, this procedure is applied to the perturbative contributions and the non-perturbative terms of mass dimensions three, four, and five. For the terms that do not possess imaginary parts, namely the non-perturbative contributions of mass dimension six, only the Fourier integral is performed and their contributions are evaluated through a direct Borel transformation.

In the final step, the same procedure applied to the physical side is adopted for the correlation function of the QCD side. The various structures of the correlation function in the theoretical side are properly ordered, the unwanted spin-$\frac{1}{2}$ contributions are removed, and the Borel transformation is applied.

The final expression of the correlation function in the QCD side, obtained after performing the aforementioned procedures, is presented in Appendix~\ref{ApenA}. This correlation function consists of two contributions: imaginary and non-imaginary parts. In this correlation function, Eq.~(\ref{QCDCorrF}), the invariant functions $\Pi_{i}^{QCD}(p^2,p'^2,q^2)$, where $i$ labels the different Lorentz structures, can be expressed in terms of double dispersion integrals as,
\begin{equation}
	\Pi_{i}^{QCD}(p^2,p'^2,q^2)= \int_{s_{min}}^{\infty}ds\int_{s'_{min}}^{\infty}ds' \frac{\rho_{i}^{QCD}(s,s',q^2)}{(s-p^2)(s'-p'^2)} + \Gamma_{i}(p^2,p'^2,q^2) \,, 
\end{equation}
where $s_{min}=(2m_s+m_b)^2$ and $s'_{min}=(2m_s+m_c)^2$. The function $\rho_{i}^{QCD}(s,s',q^2)$ represents the imaginary part of the correlation function, also referred to as the \textit{spectral density}, and is defined as $\rho_{i}^{QCD}(s,s',q^2)=\frac{1}{\pi} Im[\Pi_{i}^{QCD}(p^2,p'^2,q^2)]$, while $\Gamma_{i}(p^2,p'^2,q^2)$ denotes the non-imaginary contributions. The spectral density includes both the perturbative and non-perturbative components,
\begin{equation}
	\rho_{i}^{QCD}(s,s',q^2)= \rho_{i}^{Pert.}(s,s',q^2) + \sum_{n=3}^{5} \rho_{i}^{n}(s,s',q^2) \,, 
\end{equation}
As mentioned earlier, the non-perturbative contributions, commonly referred to as vacuum condensates, include quark condensates, gluon condensates, and quark-gluon mixed condensates.

In this step, the double Borel transformation, Eq.~(\ref{BorelT}), is first applied to suppress the contributions of the excited states and the hadronic continuum. Next, the \textit{quark–hadron duality}, a fundamental aspect of the QCD sum rule method is employed. Through the quark-hadron duality the continuum subtraction is applied on the correlation function of the theoretical side. Consequently, we have,
\begin{equation}
	\Pi_{i}^{QCD}(M^2,M'^2,s_0,s'_0,q^2)= \int_{s_{min}}^{s_0}ds\int_{s'_{min}}^{s'_0}ds' e^{-\frac{s}{M^2}} e^{-\frac{s'}{M'^2}} \rho_{i}^{QCD}(s,s',q^2) + \mathrm{\hat{B}}[\Gamma_{i}(p^2,p'^2,q^2)] \,, 
\end{equation}
where $s_0$ and $s'_0$ denote the continuum thresholds of the initial and final baryon states, respectively, arising from the application of the quark-hadron duality.

At this stage, the QCD sum rules for the form factors can be determined in terms of the baryon residues and masses, QCD parameters such as the strong coupling constant, quark condensates, gluon condensates, quark-gluon mixed condensates, quark masses and the auxiliary parameters $M^2$, $M'^2$, $s_0$, $s'_0$. This is achieved by matching the coefficients of the corresponding Lorentz structures in the correlation functions of the physical and QCD sides. In the QCD sum rule approach, the auxiliary parameters are not physical quantities. They are fixed in the numerical analysis such that the resulting physical observables exhibit minimal sensitivity to their values.

It should be stressed that three-point QCD sum rules are expected to provide reliable results only in a restricted kinematic region, typically at low to intermediate values of $q^2$, where the operator product expansion converges and the continuum contributions remain under control. In heavy-quark systems, this limitation is well known, since the interplay between the OPE and the heavy-quark mass scale may upset the naive heavy-quark power counting, leading to potential convergence issues. In the present analysis, we explicitly account for these limitations by working within carefully chosen Borel windows, determined by the standard sum-rule criteria: (i) dominance of the perturbative contribution, (ii) numerical suppression of higher-dimensional condensate contributions, and (iii) a sufficiently large pole contribution compared to the continuum. The form factors are therefore extracted only in the region where the sum rules exhibit acceptable stability with respect to the auxiliary parameters. To obtain their behavior over the full physical kinematic region, we subsequently employ an appropriate fit function, which serves as a phenomenological extrapolation rather than a direct sum-rule prediction.

\section{Numerical Calculations}
\label{NumCalc} 

In this section, we present the numerical evaluation of the QCD sum rules for the form factors. The sum rules are analyzed and the best fitting functions are determined to examine the behavior of the form factors as functions of $q^2$ over the entire physical region. In our calculations, two types of parameters are involved. The first type consists of four auxiliary parameters: the continuum thresholds $s_0$ and $s'_0$, and the Borel mass parameters $M^2$ and $M'^2$, which arise from the application of the quark-hadron duality and the Borel transformation, respectively. Since these parameters are not physical, the extracted physical quantities such as the form factors are expected to remain stable and show minimal dependence on their variations, in accordance with the principles of the QCD sum rule approach. The ranges of these parameters that satisfy this stability criterion are referred to as the \textit{working regions} of the auxiliary parameters. These working regions are determined based on the above criterion along with the following conditions and requirements.

The upper bounds of the Borel parameters, $M^2$ and $M'^2$, are determined by imposing the pole dominance condition, which requires that the contributions from the ground state exceed those from the excited states and the hadronic continuum, 
\begin{equation}
	Pole\ Contribution = \frac{\Pi^{QCD}(M^2,M'^2,s_0,s'_0)}{\Pi^{QCD}(M^2,M'^2,\infty,\infty)} \geq 0.5 \,. 
\end{equation}
The lower bounds of the Borel parameters are determined by requiring convergence of the OPE series. Specifically, the perturbative contributions should dominate over the non-perturbative ones and the higher dimensional non-perturbative terms should provide progressively smaller contributions. For this purpose, we impose the condition that the contribution of the highest-dimensional non-perturbative operator (dimension-6) remains below approximately 5\% of the total perturbative and non-perturbative contributions. Explicitly,
\begin{equation}\label{OPEConv}
	R(M^2,M'^2) = \frac{\Pi^{QCD-dim6}(M^2,M'^2,s_0,s'_0)}{\Pi^{QCD}(M^2,M'^2,s_0,s'_0)} \leq 0.05 \,. 
\end{equation}
These conditions and requirements define the working regions of the Borel parameters as follows,
\begin{equation}
	9\ \mathrm{GeV^2} \leq M^2 \leq 12\ \mathrm{GeV^2},\ \     6\ \mathrm{GeV^2} \leq M'^2 \leq 9\ \mathrm{GeV^2}  \,. 
\end{equation}
Although some form factors may exhibit relatively smooth behavior outside the adopted Borel windows, those regions are not considered reliable because either the pole dominance deteriorates or the convergence of the OPE becomes insufficient.

The working regions of the continuum thresholds, $s_0$ and $s'_0$, are chosen to suppress the contributions of the excited states and continuum in the initial and final baryon states by restricting the upper limits of the integrals in the correlation function. To determine these regions, the stability of the QCD sum rules for the form factors within the previously established Borel parameters ranges must also be verified. Following this careful selection, the working regions of the continuum thresholds are found to be,
\begin{equation} \label{contin_thresh}
	\begin{split}
		 (m_{\Omega_{b}^{*}}+0.3)^2\ \mathrm{GeV^2} \leq s_0 \leq (m_{\Omega_{b}^{*}}+0.5)^2\ \mathrm{GeV^2} \,, \\
		 (m_{\Omega_{c}^{*}}+0.3)^2\ \mathrm{GeV^2} \leq s'_0 \leq (m_{\Omega_{c}^{*}}+0.5)^2\ \mathrm{GeV^2}  \,.
	\end{split}
\end{equation}
The continuum thresholds $s_0$ and $s_0'$ are auxiliary parameters related to the energies of the first excited states in the initial and final baryonic channels. In practice, they define the upper limits of the hadronic spectral integrals in the sum rules, where the contributions above these thresholds are modeled through the quark-hadron duality ansatz and subtracted as continuum contributions. The adopted intervals in Eq. (\ref{contin_thresh}) correspond to excitation energies of about $0.3$-$0.5~\mathrm{GeV}$ above the ground-state baryon masses, which is consistent with the expected location of the first excited states in the corresponding channels. Within these working regions, the extracted form factors exhibit stable behavior under the variations of the auxiliary parameters. The same continuum-threshold intervals are used consistently for all Lorentz-structure sets considered in the present analysis.

In the numerical analysis, the form factors are evaluated for a broad set of combinations of the auxiliary parameters within the adopted working regions. The final central values and uncertainties are extracted from the corresponding stable results under variations of the Borel parameters and continuum thresholds. We have also verified that correlated variations of the two Borel parameters, motivated by the hierarchy of the initial and final baryon masses, do not lead to significant changes in the resulting form factors within the quoted uncertainties.

\begin{figure}[h!]
	\centering
	\includegraphics[width=0.45\textwidth]{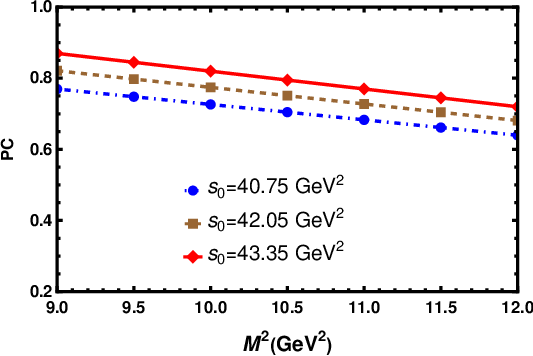}
	\includegraphics[width=0.45\textwidth]{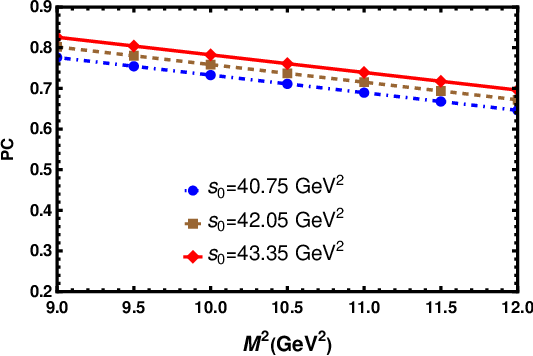}
	\caption{Pole contribution of $\Omega_b^*$ as a function of the Borel parameter $M^2$ for different values of the continuum threshold $s_0$: left panel for the Lorentz structure $p_{\rho}g_{\mu\nu}\slashed{p}\slashed{p'}$, and right panel for the Lorentz structure $p'_{\nu}g_{\mu\rho}\slashed{p}\slashed{p'}$.}
	\label{PC}
\end{figure}

\begin{figure}[h!]
	\centering
	\includegraphics[width=0.45\textwidth]{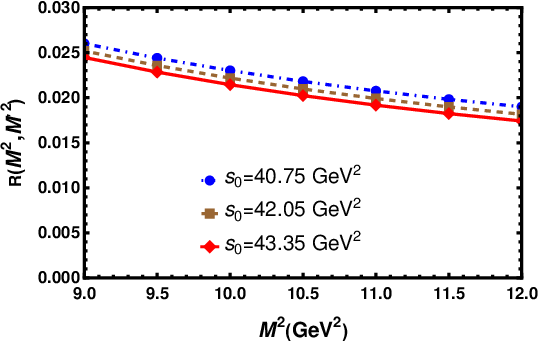}
	\includegraphics[width=0.45\textwidth]{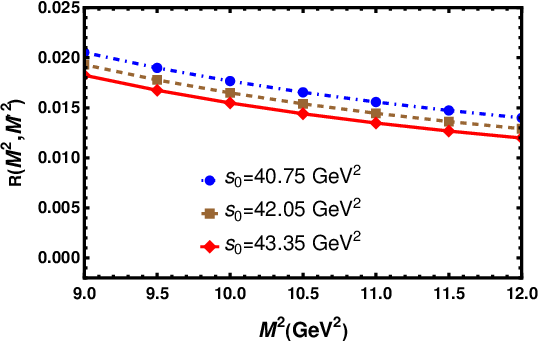}
	\caption{OPE convergence of $\Omega_b^*$ as a function of the Borel parameter $M^2$ for fixed values of the continuum threshold $s_0$: left panel for the Lorentz structure $p_{\mu}g_{\nu\rho}\slashed{p'}$, and right panel for the Lorentz structure $g_{\nu\rho}\gamma_{\mu}\slashed{p}\slashed{p'}$.}
	\label{OPE}
\end{figure}

To further verify the reliability of the adopted working regions, we also examined, as representative examples, the pole contribution and the convergence of the OPE for selected Lorentz structures from the three Lorentz-structure sets. The corresponding results are shown in Figs.~\ref{PC} and~\ref{OPE} as functions of the Borel parameter $M^2$, with $M'^2$ fixed at its central value, for different values of the continuum threshold $s_0$. Similar behavior is also observed when varying $M'^2$ and $s'_0$ while fixing $M^2$ at its central value. We confirmed that these analyses satisfy the standard reliability criteria of the QCD sum-rule method. In particular, the pole contribution remains sufficiently large within the chosen Borel windows, while the perturbative part provides the dominant contribution to the correlation function. Furthermore, the dimension-6 contributions remain below approximately 5\% of the total correlation function, consistent with Eq. (\ref{OPEConv}), indicating satisfactory convergence of the OPE and supporting the reliability of the adopted working regions of the auxiliary parameters.

Second, the parameters related to the physical and QCD sides, whose numerical values are listed in Table~\ref{InParam}, are used as input for the numerical calculations. The heavy-quark masses used in the present analysis correspond to the $\overline{\mathrm{MS}}$ scheme, namely $\bar{m}_b(\bar{m}_b)$ and $\bar{m}_c(\bar{m}_c)$. The vacuum condensates and other nonperturbative input parameters are evaluated at the conventional hadronic renormalization scale $\mu \simeq 1~\mathrm{GeV}$ commonly used in QCD sum rule analyses. In the following section, these parameters are employed to examine the stability of the form factors within the chosen working regions and to determine their optimal fitting functions, which describe the numerical behavior of the form factors as functions of $q^2$.
\begin{table}[ht]
	\centering
	\caption{Values of the second type of parameters used in numerical calculations}
	\begin{tabular}{cc} 
		\toprule
		\textbf{Parameters} & \textbf{Values}  \\
		\midrule
		$m_s$    & $(93.4_{-3.4}^{+8.6})\ $MeV \cite{ParticleDataGroup:2024cfk}     \\ 
		$m_b$    & $(4.18_{-0.02}^{+0.03})\ $GeV  \cite{ParticleDataGroup:2024cfk}  \\ 
		$m_c$    & $(1.27\pm 0.02)\ $GeV  \cite{ParticleDataGroup:2024cfk}   \\ 
		$m_e$    & $0.51\ $MeV    \cite{ParticleDataGroup:2024cfk}  \\ 
		$m_{\mu}$    & $105\ $MeV  \cite{ParticleDataGroup:2024cfk}   \\ 
		$m_{\tau}$    & $1776\ $MeV  \cite{ParticleDataGroup:2024cfk}   \\ 
		$m_{\Omega_{b}^{*}}$    & $(6084\pm84)\ $MeV \cite{Agaev:2017jyt}    \\ 
		$m_{\Omega_{c}^{*}}$    & $(2765.9\pm2)\ $MeV \cite{ParticleDataGroup:2024cfk}    \\ 
		$m_{0}^{2}$    & $(0.8\pm0.2)\ $$\mathrm{GeV}^2$  \cite{Belyaev:1982sa,Belyaev:1982cd,Ioffe:2005ym}    \\ 
		$\lambda_{\Omega_{b}^{*}}$    & $(9.3\pm1.4)\ .\ 10^{-2}\ $$\mathrm{GeV}^3$ \cite{Agaev:2017jyt}   \\ 
		$\lambda_{\Omega_{c}^{*}}$    & $(7.1\pm1.0)\ .\ 10^{-2}\ $$\mathrm{GeV}^3$  \cite{Agaev:2017jyt}  \\ 
		$G_F$    & $1.17 \times 10^{-5}\ $ $\mathrm{GeV}^{-2}$ \cite{ParticleDataGroup:2024cfk}    \\
		$V_{cb}$    & $(39\pm1.1)\ .\ 10^{-3}\ $   \cite{ParticleDataGroup:2024cfk}  \\ 
		$\langle \bar{u}u \rangle$    & $-(0.24\pm0.01)^3 $  $\mathrm{GeV}^3$  \cite{Belyaev:1982sa,Belyaev:1982cd}   \\ 
		$\langle \bar{s}s \rangle$    & $(0.8\pm0.1) \langle \bar{u}u \rangle $ $\mathrm{GeV}^3$  \cite{Belyaev:1982sa,Belyaev:1982cd}     \\ 
		$\braket{0| \frac{1}{\pi} \alpha_s G^2 |0}$    & $(0.012\pm0.004)\ $ $\mathrm{GeV}^4$  \cite{Belyaev:1982sa,Belyaev:1982cd,Ioffe:2005ym} \\
		\bottomrule
	\end{tabular}
	\label{InParam}
\end{table}	
\subsection{Form Factors}
\begin{figure}[h!]
	\centering
	\includegraphics[width=0.32\textwidth]{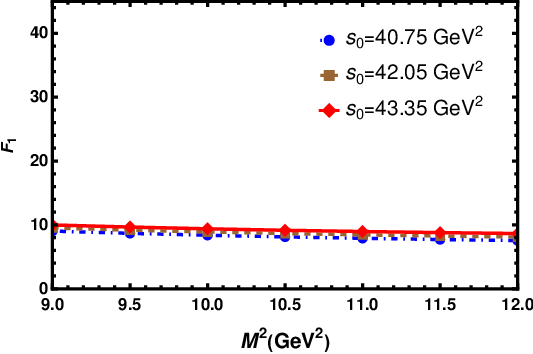}
	\includegraphics[width=0.32\textwidth]{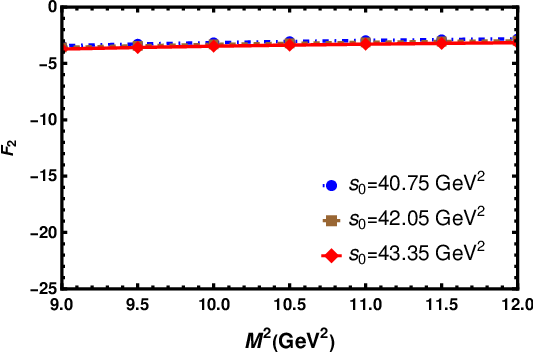}
	\includegraphics[width=0.32\textwidth]{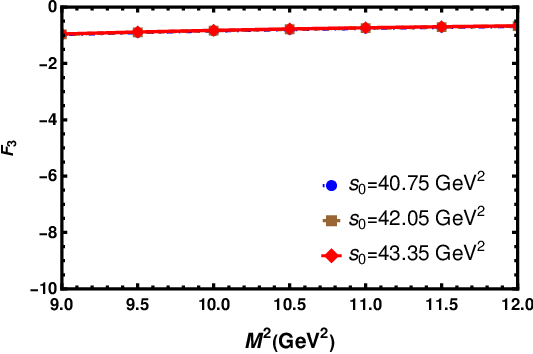}
	\includegraphics[width=0.32\textwidth]{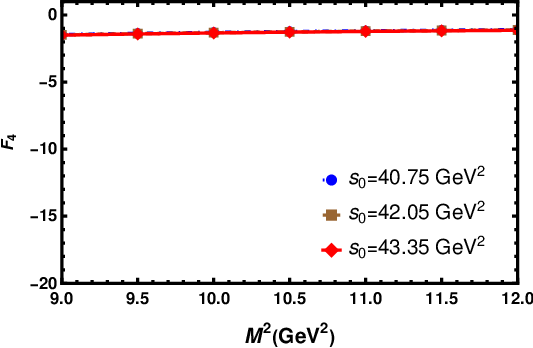}
	\includegraphics[width=0.32\textwidth]{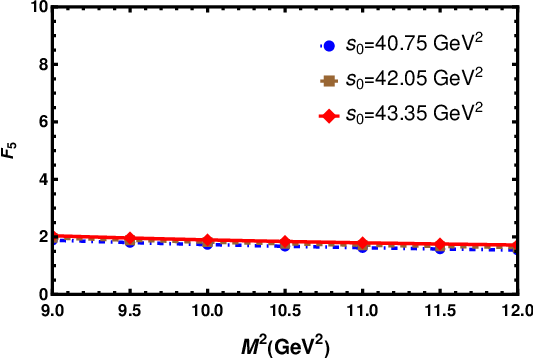}
	\includegraphics[width=0.32\textwidth]{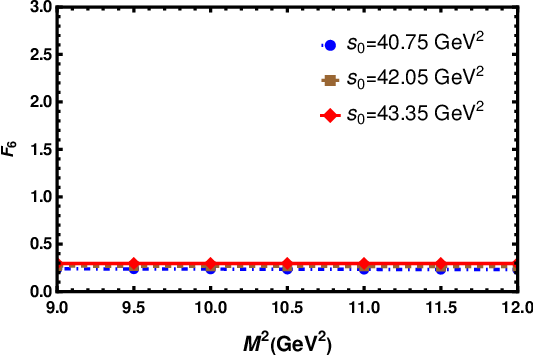}
	\includegraphics[width=0.32\textwidth]{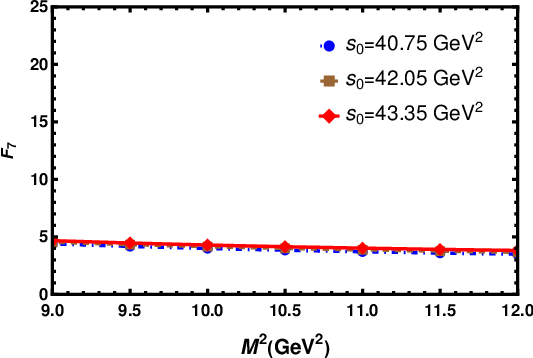}
	\includegraphics[width=0.32\textwidth]{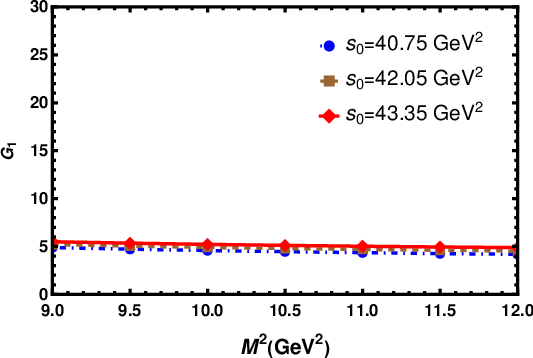}
	\includegraphics[width=0.32\textwidth]{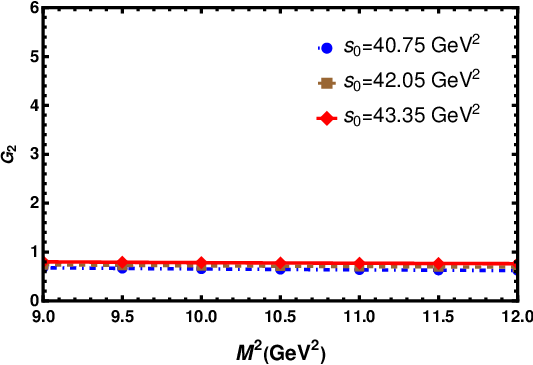}
	\includegraphics[width=0.32\textwidth]{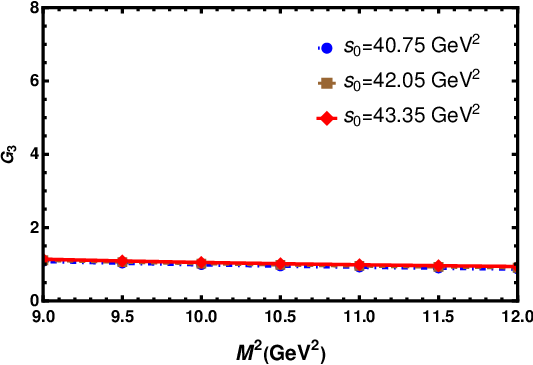}
	\includegraphics[width=0.32\textwidth]{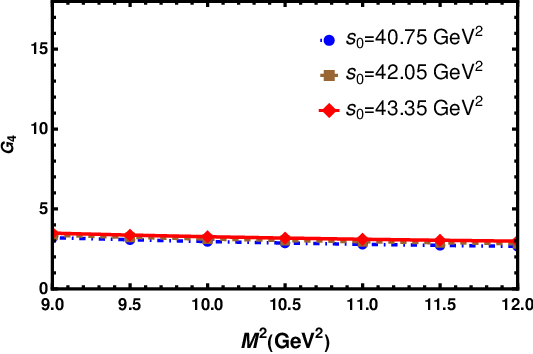}
	\includegraphics[width=0.32\textwidth]{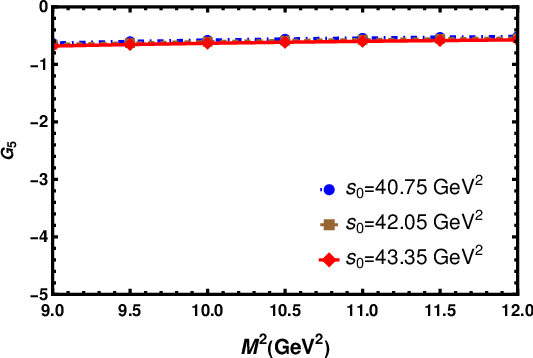}
	\includegraphics[width=0.32\textwidth]{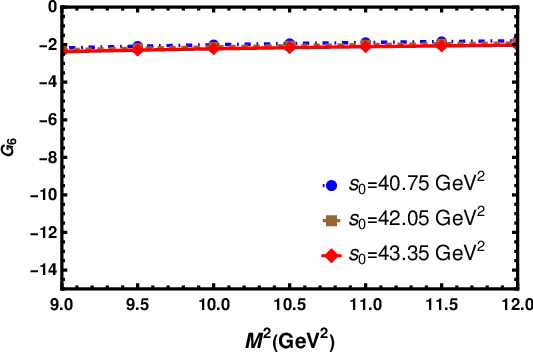}
	\includegraphics[width=0.32\textwidth]{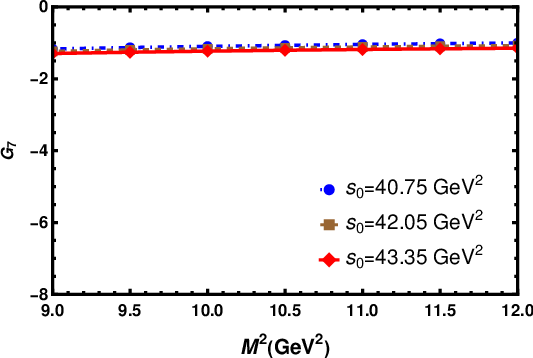}
	\caption{Dependence of the form factors, $F_i$ and $G_i$, on the auxiliary parameters $M^2$ and $s_0$ at $q^2 = 0$, with the other auxiliary parameters fixed at their central values, for the first set of selected structures.}
	\label{FFM1s0}
\end{figure}

\begin{figure}[h!]
	\centering
	\includegraphics[width=0.32\textwidth]{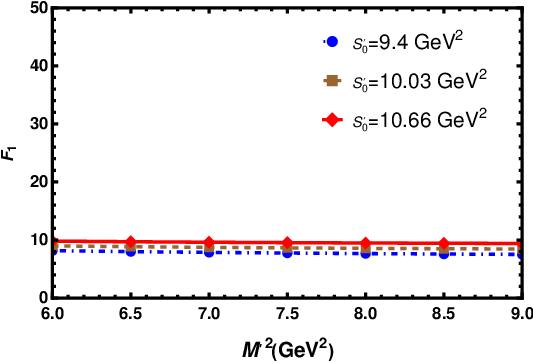}
	\includegraphics[width=0.32\textwidth]{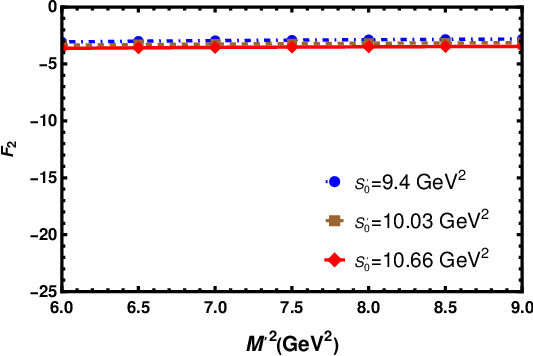}
	\includegraphics[width=0.32\textwidth]{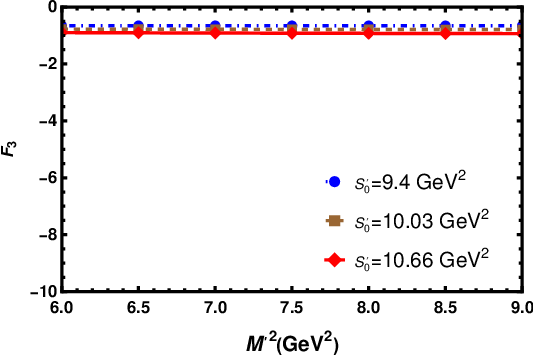}
	\includegraphics[width=0.32\textwidth]{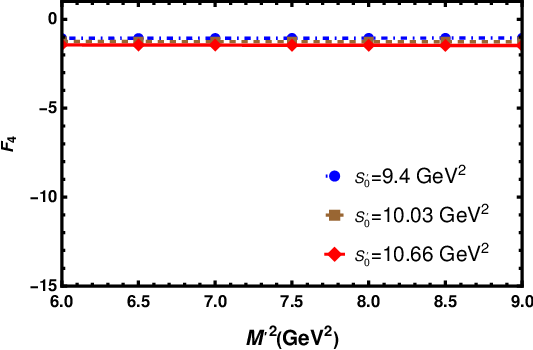}
	\includegraphics[width=0.32\textwidth]{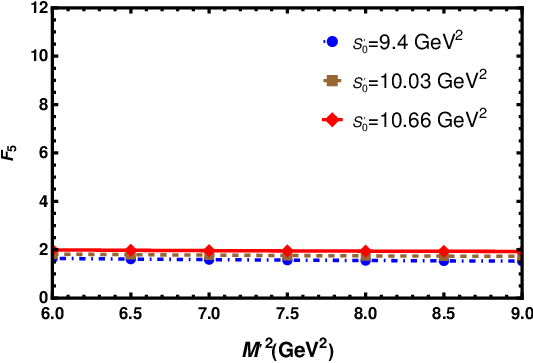}
	\includegraphics[width=0.32\textwidth]{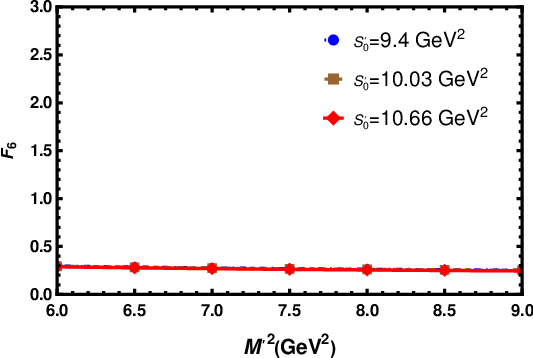}
	\includegraphics[width=0.32\textwidth]{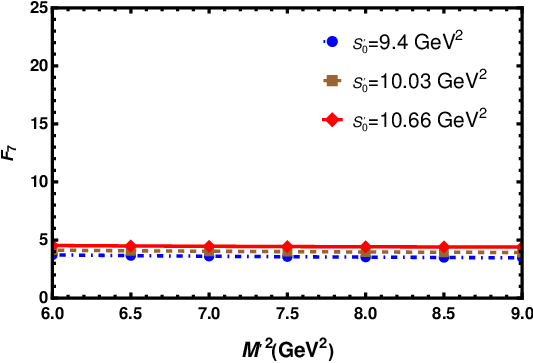}
	\includegraphics[width=0.32\textwidth]{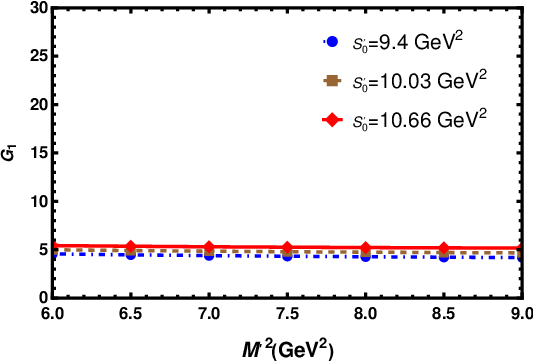}
	\includegraphics[width=0.32\textwidth]{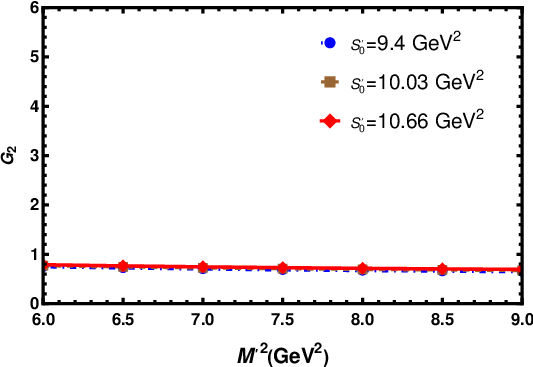}
	\includegraphics[width=0.32\textwidth]{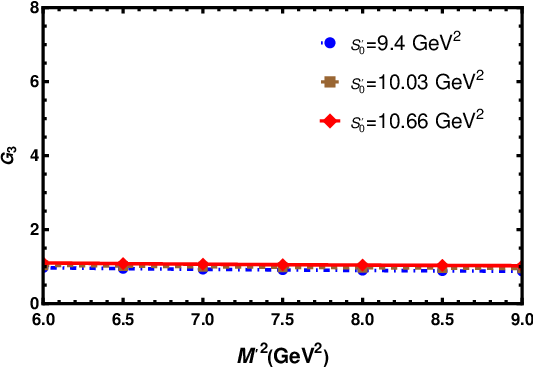}
	\includegraphics[width=0.32\textwidth]{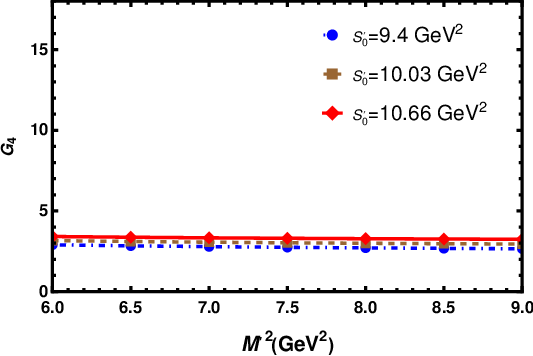}
	\includegraphics[width=0.32\textwidth]{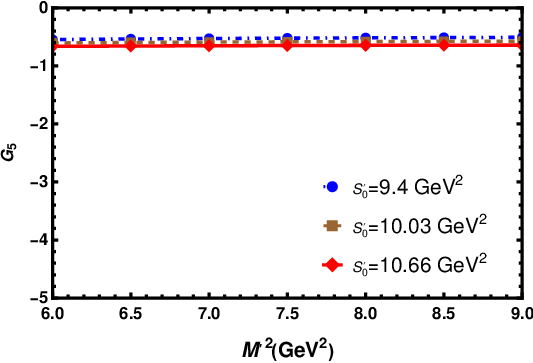}
	\includegraphics[width=0.32\textwidth]{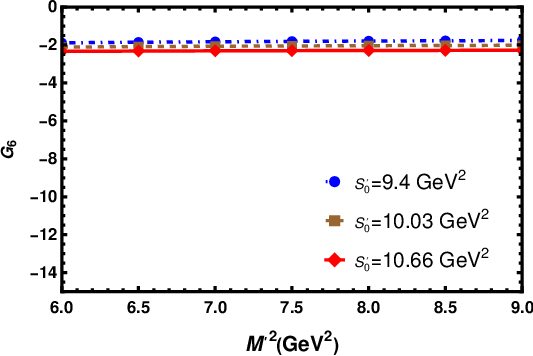}
	\includegraphics[width=0.32\textwidth]{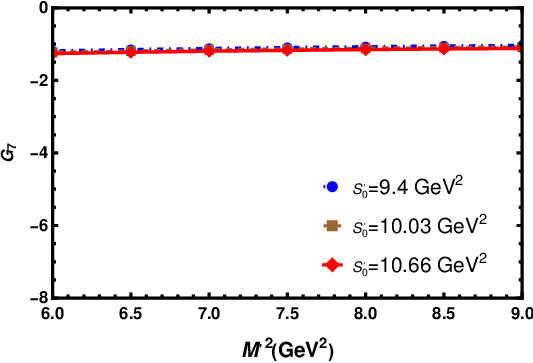}
	\caption{Dependence of the form factors, $F_i$ and $G_i$, on the auxiliary parameters $M'^2$ and $s'_0$ at $q^2 = 0$, with the other auxiliary parameters fixed at their central values, for the first set of selected structures.}
	\label{FFM1sp0}
\end{figure}

As mentioned above, the $\tfrac{3}{2}\to\tfrac{3}{2}$ weak transition is described by fourteen independent form factors, denoted by $F_i$ and $G_i$. In the numerical analysis, we first examine the dependence of these form factors on the auxiliary parameters of the QCD sum rules. In particular, their behavior within the working regions of the Borel mass parameters $M^2$ and $M'^2$, as well as the continuum thresholds $s_0$ and $s'_0$, is illustrated in Figs.~\ref{FFM1s0} and~\ref{FFM1sp0} for representative Lorentz structures.
The obtained form factors display good stability, smooth behavior, and only mild sensitivity to variations of the auxiliary parameters within the selected working intervals. This stability indicates that the adopted ranges of $M^2$, $M'^2$, $s_0$, and $s'_0$ satisfy the standard QCD sum-rule requirements and that the extracted form factors are reliable within these domains. In particular, the lower bounds of the Borel windows are fixed by demanding a satisfactory convergence of the operator product expansion (OPE), such that contributions from higher-dimensional condensates remain under control, while the upper bounds are chosen to ensure a sufficient suppression of excited and continuum states relative to the ground-state contribution.
Consequently, the working regions are determined by balancing OPE convergence against pole dominance, leading to a stable extraction of the form factors with respect to the auxiliary parameters. This procedure ensures that the resulting form factors provide a consistent and trustworthy input for the subsequent phenomenological analysis of the semileptonic decay observables. Similar stability behavior is also observed for the form factors of second and third Lorentz-structure sets.

Next, we analyze the numerical behavior of the form factors as functions of the momentum transfer squared, $q^2$. It is well known that QCD sum rules yield reliable predictions for form factors only within a restricted region of $q^2$, typically near the spacelike or low-$q^2$ domain, where the operator product expansion converges and the continuum contributions remain under control. In order to extrapolate the form factors to the entire physical region, $m_{\ell}^2\leq q^2 \leq (m_{\Omega_b^*}-m_{\Omega_c^*})^2$, a suitable parametrization is required. For this purpose, we adopt a phenomenological polynomial fit function, which provides a simple and efficient description of the $q^2$ dependence and captures the local curvature of the sum-rule results near $q^2=0$ through a limited number of free parameters. The $q^2$ dependence of the form factors $F_i$ and $G_i$ is parameterized as
\begin{equation}
	\mathcal{F}(q^2) = \frac{\mathcal{F}(0)}{1-a\ (\frac{q^2}{m_{\Omega_{b}^{*}}^{2}}) +b\ (\frac{q^2}{m_{\Omega_{b}^{*}}^{2}})^2+c\ (\frac{q^2}{m_{\Omega_{b}^{*}}^{2}})^3+d\ (\frac{q^2}{m_{\Omega_{b}^{*}}^{2}})^4} \,. 
\end{equation}
The fitting functions are determined separately for the form factors associated with three different sets of selected Lorentz structures. The corresponding fit parameters, $\mathcal{F}(0)$, $a$, $b$, $c$, and $d$, extracted using the central values of the auxiliary parameters, are presented in Tables~\ref{VecF} and~\ref{AVecF} for the first set of structures, Tables~\ref{VecFset2} and~\ref{AVecFset2} for the second set of structures, and Tables~\ref{VecFset3} and~\ref{AVecFset3} for the third set of structures.

It should be emphasized that this parametrization is not unique, and alternative functional forms and fitting strategies can also be employed. A theoretically well-motivated alternative is the $z$-expansion~\cite{Cui:2022zwm}, which exploits the analytic properties of the form factors through a conformal mapping based on the lightest physical branch point. This approach ensures controlled convergence, reduces model dependence, and is widely used in lattice-QCD analyses. However, in QCD sum-rule calculations the form factors are available only at a finite number of points within a limited kinematic region. In this context, the use of a simple pole-like or polynomial parametrization provides a stable and efficient representation of the sum-rule results with a minimal number of parameters, and has been extensively adopted in the literature on heavy-baryon transitions~\cite{Najjar:2024deh,Neishabouri:2024gbc,Tousi:2024usi,Khajouei:2024frw}. In the present work, the polynomial fit serves primarily as an interpolation and mild extrapolation tool to smoothly connect the reliable sum-rule predictions across the physical $q^2$ region. We stress that the physical observables are predominantly determined by the kinematic region where the QCD sum-rule results are trustworthy, and that the uncertainties associated with the choice of parametrization are included in the quoted error estimates. Consequently, the use of alternative parametrizations, such as the $z$-expansion, is not expected to lead to significant numerical differences within the present theoretical accuracy.
\begin{table}[ht]
	\centering
	\caption{Parameters of the fitting functions for the form factors $F_i$ corresponding to the first set of selected structures}
	\resizebox{\textwidth}{!}{%
	\begin{tabular}{cccccccc} 
		\toprule
		\textbf{Set 1} & \textbf{$F_1(q^2)$} & \textbf{$F_2(q^2)$} & \textbf{$F_3(q^2)$} & \textbf{$F_4(q^2)$} & \textbf{$F_5(q^2)$} & \textbf{$F_6(q^2)$} & \textbf{$F_7(q^2)$} \\
		\midrule
		$\mathcal{F}(q^2=0)$ & $8.73\pm0.96$    & $-3.25\pm0.36$  & $-0.80\pm0.14$    & $-1.28\pm0.19$ & $1.78\pm0.20$  & $0.27\pm0.03$    & $4.05\pm0.48$ \\
		a & $1.48$    & $1.61$  & $2.11$    & $2.23$   & $2.24$  & $2.07$  & $1.70$  \\
		b & $0.36$    & $0.44$  & $1.31$    & $1.34$  & $1.68$  & $1.94$   & $0.57$ \\
		c & $0.06$    & $0.14$  & $-0.21$    & $0.26$  & $-0.43$  & $-0.96$    & $0.13$ \\
		d & $0.01$    & $0.004$  & $-0.02$    & $-0.41$  & $-0.01$  & $1.05$  & $0.02$ \\
		\bottomrule
	\end{tabular}
}
	\label{VecF}
\end{table}
\begin{table}[ht]
	\centering
	\caption{Parameters of the fitting functions for the form factors $G_i$ corresponding to the first set of selected structures}
	\resizebox{\textwidth}{!}{%
	\begin{tabular}{cccccccc} 
		\toprule
		\textbf{Set 1} & \textbf{$G_1(q^2)$} & \textbf{$G_2(q^2)$} & \textbf{$G_3(q^2)$} & \textbf{$G_4(q^2)$} & \textbf{$G_5(q^2)$} & \textbf{$G_6(q^2)$} & \textbf{$G_7(q^2)$} \\
		\midrule
		$\mathcal{F}(q^2=0)$ & $4.85\pm0.52$    & $0.72\pm0.07$  & $0.99\pm0.10$    & $3.07\pm0.31$ & $-0.59\pm0.06$  & $-2.06\pm0.23$    & $-1.15\pm0.11$ \\
		a & $1.20$    & $1.52$  & $1.97$    & $1.79$   & $2.24$  & $2.55$  & $1.62$  \\
		b & $0.20$    & $0.48$  & $1.21$    & $0.95$  & $1.69$  & $2.15$   & $0.73$ \\
		c & $0.04$    & $0.05$  & $-0.32$    & $-0.11$  & $-0.43$  & $-0.60$    & $-0.08$ \\
		d & $-0.004$    & $-0.01$  & $0.05$    & $-0.02$  & $-0.01$  & $-0.02$  & $-0.008$ \\
		\bottomrule
	\end{tabular}
}
	\label{AVecF}
\end{table}
\begin{table}[ht]
	\centering
	\caption{Parameters of the fitting functions for the form factors $F_i$ corresponding to the second set of selected structures}
	\resizebox{\textwidth}{!}{%
		\begin{tabular}{cccccccc} 
			\toprule
			\textbf{Set 2} & \textbf{$F_1(q^2)$} & \textbf{$F_2(q^2)$} & \textbf{$F_3(q^2)$} & \textbf{$F_4(q^2)$} & \textbf{$F_5(q^2)$} & \textbf{$F_6(q^2)$} & \textbf{$F_7(q^2)$} \\
			\midrule
			$\mathcal{F}(q^2=0)$ & $9.20\pm1.05$ & $-3.47\pm0.44$ & $-0.60\pm0.05$ & $-3.27\pm0.32$ & $2.91\pm0.29$  & $-0.85\pm0.09$    & $2.46\pm0.20$ \\
			a & $1.65$    & $1.73$  & $1.56$    & $1.86$  & $2.95$  & $2.47$  & $1.44$  \\
			b & $0.56$    & $0.58$  & $0.51$    & $0.69$  & $1.69$  & $1.70$   & $0.26$ \\
			c & $0.005$   & $0.14$  & $0.09$   & $0.32$  & $-0.38$ & $-0.02$    & $0.11$ \\
			d & $0.03$    & $-0.01$ & $-0.05$   & $-0.12$ & $-0.04$ & $-0.2$  & $0.03$ \\
			\bottomrule
		\end{tabular}
	}
	\label{VecFset2}
\end{table}
\begin{table}[ht]
	\centering
	\caption{Parameters of the fitting functions for the form factors $G_i$ corresponding to the second set of selected structures}
	\resizebox{\textwidth}{!}{%
		\begin{tabular}{cccccccc} 
			\toprule
			\textbf{Set 2} & \textbf{$G_1(q^2)$} & \textbf{$G_2(q^2)$} & \textbf{$G_3(q^2)$} & \textbf{$G_4(q^2)$} & \textbf{$G_5(q^2)$} & \textbf{$G_6(q^2)$} & \textbf{$G_7(q^2)$} \\
			\midrule
			$\mathcal{F}(q^2=0)$ & $3.07\pm0.29$    & $-2.63\pm0.28$  & $-2.34\pm0.27$    & $3.41\pm0.33$ & $0.46\pm0.05$  & $-3.15\pm0.33$    & $-0.27\pm0.05$ \\
			a & $0.87$    & $1.82$  & $1.67$    & $1.70$   & $1.96$  & $2.41$  & $1.06$  \\
			b & $0.17$    & $0.85$  & $0.58$    & $0.78$  & $0.92$  & $1.83$   & $-0.31$ \\
			c & $0.01$    & $-0.01$  & $0.09$    & $-0.01$  & $0.02$  & $-0.39$    & $-0.13$ \\
			d & $0.27$    & $-0.04$  & $-0.03$    & $-0.04$  & $0.01$  & $-0.05$  & $0.44$ \\
			\bottomrule
		\end{tabular}
	}
	\label{AVecFset2}
\end{table}
\begin{table}[ht]
	\centering
	\caption{Parameters of the fitting functions for the form factors $F_i$ corresponding to the third set of selected structures}
	\resizebox{\textwidth}{!}{%
		\begin{tabular}{cccccccc} 
			\toprule
			\textbf{Set 3} & \textbf{$F_1(q^2)$} & \textbf{$F_2(q^2)$} & \textbf{$F_3(q^2)$} & \textbf{$F_4(q^2)$} & \textbf{$F_5(q^2)$} & \textbf{$F_6(q^2)$} & \textbf{$F_7(q^2)$} \\
			\midrule
			$\mathcal{F}(q^2=0)$ & $6.76\pm0.86$    & $-1.69\pm0.16$  & $1.19\pm0.16$    & $-2.57\pm0.26$ & $1.78\pm0.20$  & $0.27\pm0.03$    & $5.41\pm0.67$ \\
			a & $1.91$    & $2.25$  & $1.07$    & $1.26$   & $2.24$  & $2.07$  & $1.86$  \\
			b & $1.00$    & $1.47$  & $0.19$    & $0.19$  & $1.68$  & $1.94$   & $0.82$ \\
			c & $-0.20$    & $-0.11$  & $0.03$    & $0.15$  & $-0.43$  & $-0.96$    & $0.09$ \\
			d & $0.08$    & $-0.12$  & $0.04$    & $0.23$  & $-0.01$  & $1.05$  & $-0.07$ \\
			\bottomrule
		\end{tabular}
	}
	\label{VecFset3}
\end{table}
\begin{table}[ht]
	\centering
	\caption{Parameters of the fitting functions for the form factors $G_i$ corresponding to the third set of selected structures}
	\resizebox{\textwidth}{!}{%
		\begin{tabular}{cccccccc} 
			\toprule
			\textbf{Set 3} & \textbf{$G_1(q^2)$} & \textbf{$G_2(q^2)$} & \textbf{$G_3(q^2)$} & \textbf{$G_4(q^2)$} & \textbf{$G_5(q^2)$} & \textbf{$G_6(q^2)$} & \textbf{$G_7(q^2)$} \\
			\midrule
			$\mathcal{F}(q^2=0)$ & $-1.65\pm0.19$    & $-2.92\pm0.29$  & $-2.09\pm0.25$    & $2.31\pm0.24$ & $0.46\pm0.05$  & $-3.15\pm0.33$    & $0.97\pm0.09$ \\
			a & $1.46$    & $1.52$  & $2.07$    & $1.73$   & $1.96$  & $2.41$  & $1.76$  \\
			b & $0.34$    & $0.45$  & $1.27$    & $0.76$  & $0.92$  & $1.83$   & $0.87$ \\
			c & $0.13$    & $0.08$  & $-0.21$    & $0.04$  & $0.02$  & $-0.39$    & $-0.11$ \\
			d & $-0.01$    & $-0.01$  & $-0.01$    & $-0.06$  & $0.01$  & $-0.06$  & $-0.001$ \\
			\bottomrule
		\end{tabular}
	}
	\label{AVecFset3}
\end{table}

\begin{figure}[h!]
	\centering
	\includegraphics[width=0.32\textwidth]{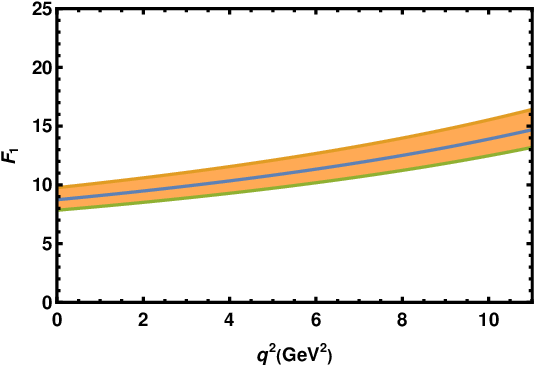}
	\includegraphics[width=0.32\textwidth]{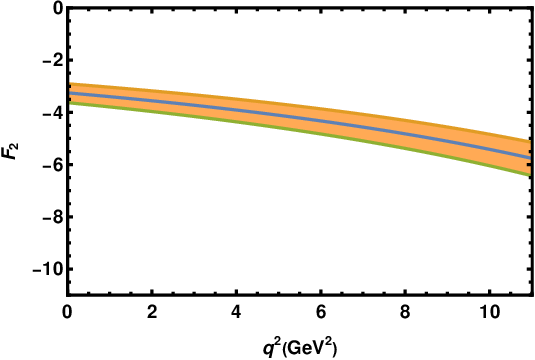}
	\includegraphics[width=0.32\textwidth]{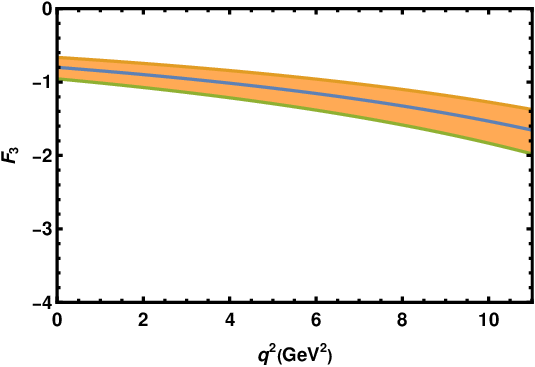}
	\includegraphics[width=0.32\textwidth]{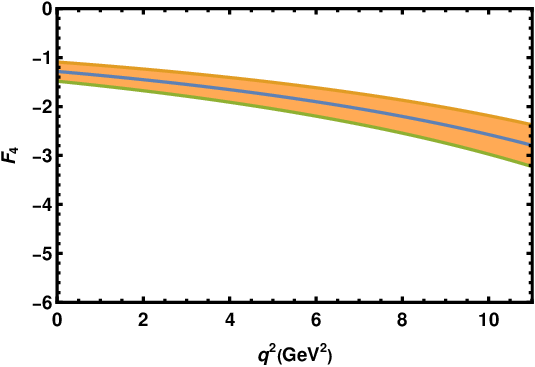}
	\includegraphics[width=0.32\textwidth]{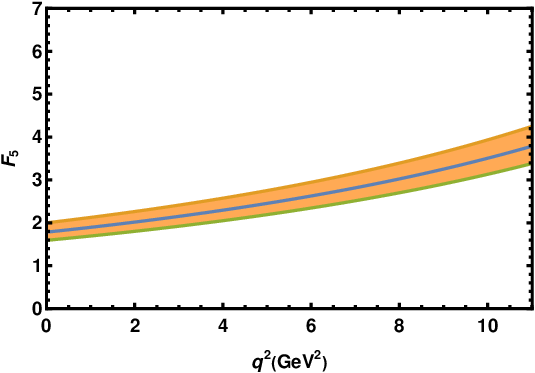}
	\includegraphics[width=0.32\textwidth]{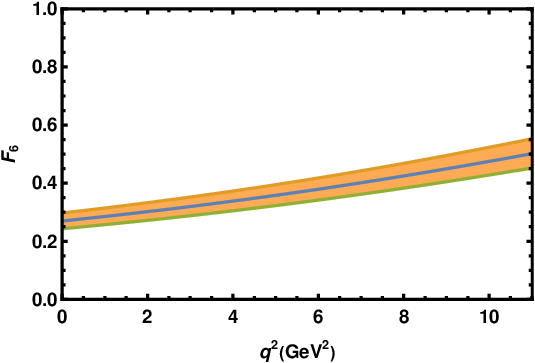}
	\includegraphics[width=0.32\textwidth]{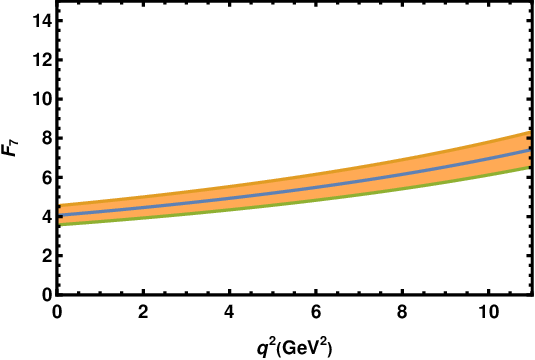}
	\includegraphics[width=0.32\textwidth]{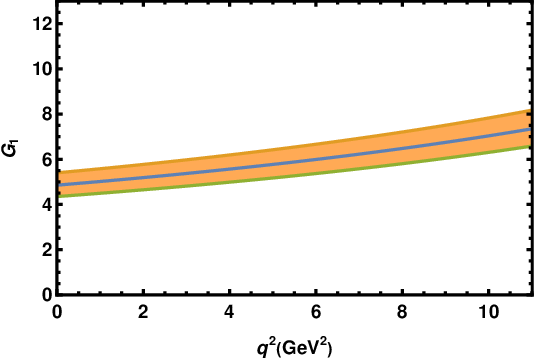}
	\includegraphics[width=0.32\textwidth]{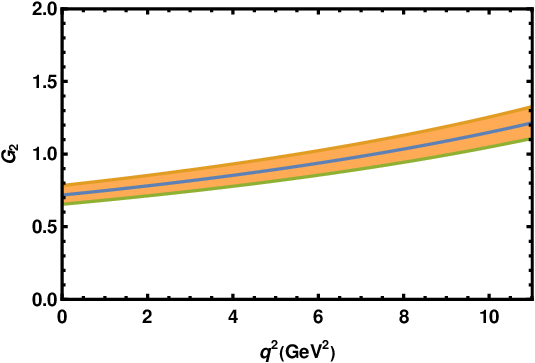}
	\includegraphics[width=0.32\textwidth]{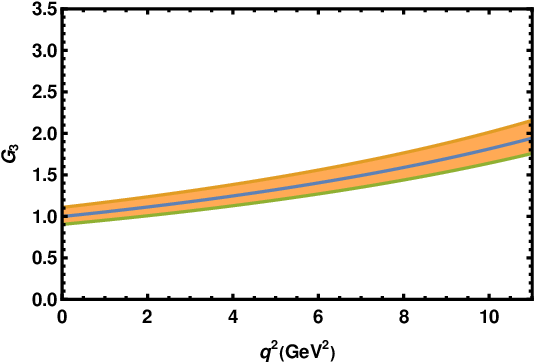}
	\includegraphics[width=0.32\textwidth]{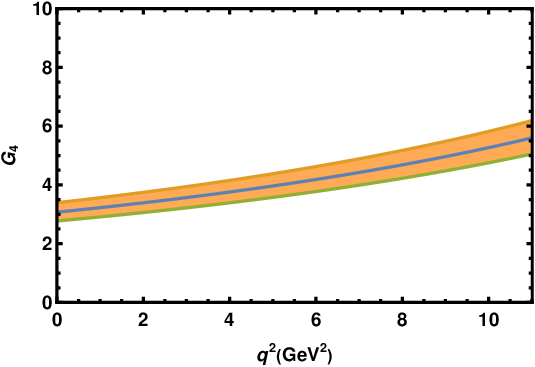}
	\includegraphics[width=0.32\textwidth]{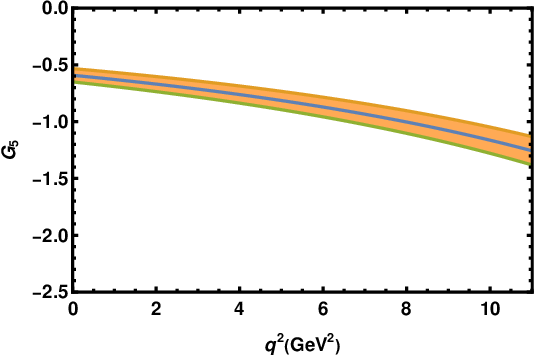}
	\includegraphics[width=0.32\textwidth]{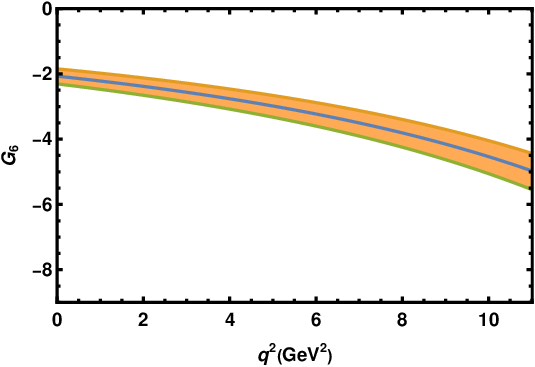}
	\includegraphics[width=0.32\textwidth]{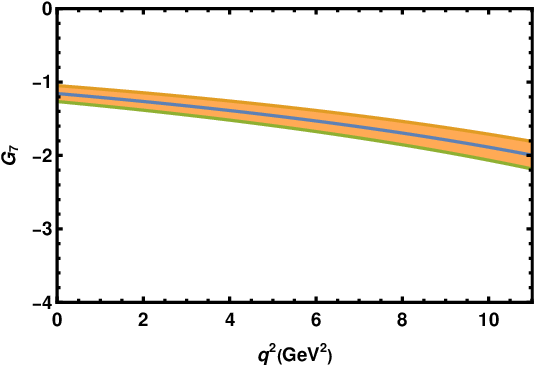}
	\caption{{The behavior of the form factors, $F_i$ and $G_i$, as functions of $q^2$  at the central values of the auxiliary parameters, with their corresponding errors from Tables~\ref{VecF} and~\ref{AVecF}.}}
	\label{ErFFq2}
\end{figure}

In this study, numerical calculations were performed on three different sets of structures to evaluate the behavior of the fourteen form factors, $F_i$ and $G_i$, of the $\Omega_{b}^{*}\rightarrow\Omega_{c}^{*} \ell \bar{\nu}_{\ell}$ weak transition as functions of $q^2$. In the QCD sum rule approach, the choice of Lorentz structures is crucial, as appropriate selections minimize uncertainties in the form factors, and more generally, in the results. For the first set, structures containing the most momenta were chosen: $g_{\nu\rho}\gamma_{\mu}\slashed{p'}$, $g_{\nu\rho}\gamma_{\mu}\slashed{p}\slashed{p'}$, $p_{\mu}g_{\nu\rho}\slashed{p}\slashed{p'}$, $p'_{\mu}g_{\nu\rho}\slashed{p}\slashed{p'}$, $p_{\rho}p'_{\nu}\gamma_{\mu}\slashed{p'}$, $p_{\rho}p'_{\nu}\gamma_{\mu}\slashed{p}\slashed{p'}$, $p_{\mu}p_{\rho}p'_{\nu}\slashed{p'}$, $p_{\rho}p'_{\mu}p'_{\nu}\slashed{p'}$, $p'_{\nu}g_{\mu\rho}\slashed{p}\slashed{p'}$, $g_{\nu\rho}\gamma_{\mu}\slashed{p'}\gamma_{5}$, $g_{\nu\rho}\gamma_{\mu}\slashed{p}\slashed{p'}\gamma_{5}$, $p_{\mu}g_{\nu\rho}\slashed{p}\slashed{p'}\gamma_{5}$, $p'_{\mu}g_{\nu\rho}\slashed{p}\slashed{p'}\gamma_{5}$, $p_{\rho}p'_{\nu}\gamma_{\mu}\slashed{p'}\gamma_{5}$, $p_{\rho}p'_{\nu}\gamma_{\mu}\slashed{p}\slashed{p'}\gamma_{5}$, $p_{\mu}p_{\rho}p'_{\nu}\slashed{p'}\gamma_{5}$, $p_{\rho}p'_{\mu}p'_{\nu}\slashed{p'}\gamma_{5}$ and $p'_{\nu}g_{\mu\rho}\slashed{p}\slashed{p'}\gamma_{5}$. The parameters of the fit functions corresponding to these structures are presented in Tables~\ref{VecF} and~\ref{AVecF}. For the second set, the chosen structures are: $g_{\nu\rho}\gamma_{\mu}\slashed{p}$, $g_{\nu\rho}\gamma_{\mu}\slashed{p}\slashed{p'}$, $p_{\mu}g_{\nu\rho}\slashed{p'}$, $p'_{\mu}g_{\nu\rho}\slashed{p}\slashed{p'}$, $p_{\rho}p'_{\nu}\gamma_{\mu}\slashed{p}$, $p_{\rho}p'_{\nu}\gamma_{\mu}\slashed{p}\slashed{p'}$, $p_{\mu}p_{\rho}p'_{\nu}\slashed{p}$, $p_{\rho}p'_{\mu}p'_{\nu}\slashed{p'}$, $p_{\rho}g_{\mu\nu}\slashed{p}\slashed{p'}$, $g_{\nu\rho}\gamma_{\mu}\slashed{p}\gamma_{5}$, $g_{\nu\rho}\gamma_{\mu}\slashed{p}\slashed{p'}\gamma_{5}$, $p'_{\mu}g_{\nu\rho}\slashed{p'}\gamma_{5}$, $p_{\mu}g_{\nu\rho}\slashed{p}\slashed{p'}\gamma_{5}$, $p_{\rho}p'_{\nu}\gamma_{\mu}\slashed{p}\gamma_{5}$, $p_{\rho}p'_{\nu}\gamma_{\mu}\slashed{p}\slashed{p'}\gamma_{5}$, $p_{\mu}p_{\rho}p'_{\nu}\slashed{p}\gamma_{5}$, $p_{\rho}p'_{\mu}p'_{\nu}\slashed{p'}\gamma_{5}$ and $p_{\rho}g_{\mu\nu}\slashed{p}\slashed{p'}\gamma_{5}$. For the third set, the selected structures include: $g_{\nu\rho}\gamma_{\mu}\slashed{p}$, $g_{\nu\rho}\gamma_{\mu}\slashed{p'}$, $p_{\mu}g_{\nu\rho}\slashed{p'}$, $p'_{\mu}g_{\nu\rho}\slashed{p'}$, $p_{\rho}p'_{\nu}\gamma_{\mu}\slashed{p}$, $p_{\rho}p'_{\nu}\gamma_{\mu}\slashed{p'}$, $p_{\mu}p_{\rho}p'_{\nu}\slashed{p'}$, $p_{\rho}p'_{\mu}p'_{\nu}\slashed{p'}$, $p'_{\nu}g_{\mu\rho}\slashed{p'}$,  $g_{\nu\rho}\gamma_{\mu}\slashed{p}\gamma_{5}$, $g_{\nu\rho}\gamma_{\mu}\slashed{p'}\gamma_{5}$,  $p_{\mu}g_{\nu\rho}\slashed{p'}\gamma_{5}$, $p'_{\mu}g_{\nu\rho}\slashed{p'}\gamma_{5}$, $p_{\rho}p'_{\nu}\gamma_{\mu}\slashed{p}\gamma_{5}$, $p_{\rho}p'_{\nu}\gamma_{\mu}\slashed{p'}\gamma_{5}$, $p_{\mu}p_{\rho}p'_{\nu}\slashed{p}\gamma_{5}$, $p_{\rho}p'_{\mu}p'_{\nu}\slashed{p'}\gamma_{5}$ and $p'_{\nu}g_{\mu\rho}\slashed{p'}\gamma_{5}$. The reported uncertainties of the form factors arise from the determination of the working regions of the auxiliary parameters and from errors in the other input parameters.  In the numerical evaluations, several systems of two equations with two unknowns were solved to extract the form factors. Fig.~\ref{ErFFq2} illustrates the numerical results of the behavior of the form factors as functions of $q^2$ including the uncertainties, for the first set of structures. 
 As expected from general kinematic considerations and heavy-to-heavy transition dynamics, all form factors exhibit a monotonically increasing behavior with increasing $q^2$. In this analysis, we employed the central values of the auxiliary parameters $M^2$, $M'^2$, $s_0$, and $s'_0$. 

This section has provided a detailed numerical analysis of the form factors. The obtained results, particularly the fitted functions of the form factors, are then employed in the next section to calculate the decay widths of the $\Omega_{b}^{*}\rightarrow\Omega_{c}^{*} \ell \bar{\nu}_{\ell}$ weak transition for all leptonic channels.
\subsection{Decay Width}
In this section, we evaluate the decay widths of the weak decay channels corresponding to the $\Omega_{b}^{*}\rightarrow\Omega_{c}^{*} \ell \bar{\nu}_{\ell}$ transition. As shown in Eqs.~(\ref{EffH}) and~(\ref{AmpT}), these weak decays proceed through both the vector and axial-vector transitions. The decay widths are calculated using the helicity amplitudes, $H_{\lambda'\lambda_{W}}$, which are defined separately for the vector and axial-vector parts as $H_{\lambda'\lambda_{W}}^{V, A}$. Here, $\lambda'=\pm1/2,\ \pm3/2$ and $\lambda_{W}=t,\ \pm1,\ 0$ denote the helicities of the final baryon state and the off-shell $W$-boson, respectively. The helicity of the initial baryon state, $\lambda$, is fixed through the relation $\lambda=\lambda'-\lambda_{W}$. The vector and axial-vector components of the helicity amplitudes are obtained from the following expressions,
\begin{equation}
	H_{\lambda'\lambda_{W}}^{V, A} =\mathcal{M}_{\mu}^{V, A}\ \bar{\epsilon}^{\ast\mu}(\lambda_{W})  \,, 
\end{equation}
where the matrix elements of the vector and axial vector transitions $\mathcal{M}_{\mu}^{V, A}$ are given by Eq.~(\ref{AmTr}) and $\bar{\epsilon}^{\ast\mu}$ denotes the polarization vector of the off-shell $W$-boson. As can be seen, the helicity amplitudes of the vector and axial-vector transitions, $H_{\lambda'\lambda_{W}}^{V, A}$, are expressed in terms of the vector form factors, $F_i$, and the axial vector form factors, $G_i$. The components of the vector and axial vector helicity amplitudes for the $\frac{3}{2}^{+}\rightarrow\frac{3}{2}^{+}$ transitions are defined by \cite{Faessler:2009xn},    
\begin{align}
		H_{\frac{1}{2}t}^{V} =& -\frac{1+2\omega}{3} H_{\frac{1}{2}t}^{V}(F_1,F_2,F_3) + \frac{2}{3}(\omega^2-1)\frac{m_{\Omega_{c}^{*}}}{m_{\Omega_{b}^{*}}} H_{\frac{1}{2}t}^{V} (F_4,F_5,F_6)  \,, \nonumber \\
		H_{\frac{1}{2}0}^{V} =& -\frac{1+2\omega}{3} H_{\frac{1}{2}0}^{V}(F_1,F_2,F_3) + \frac{2}{3}(\omega^2-1)\frac{m_{\Omega_{c}^{*}}}{m_{\Omega_{b}^{*}}} H_{\frac{1}{2}0}^{V}(F_4,F_5,F_6) - \frac{1}{3} \alpha_{\frac{1}{2}1}^{A}(\omega+1) \frac{\sqrt{2q^2}}{m_{\Omega_{b}^{*}}}G_7   \,, \nonumber \\ 
		H_{\frac{1}{2}1}^{V} =& -\frac{2\omega}{3} H_{\frac{1}{2}1}^{V}(F_1,F_2,F_3) + \frac{2}{3}(\omega^2-1)\frac{m_{\Omega_{c}^{*}}}{m_{\Omega_{b}^{*}}} H_{\frac{1}{2}1}^{V}(F_4,F_5,F_6) + \frac{1}{3} \alpha_{\frac{1}{2}1}^{V}(\omega+1) \frac{m_{+}}{m_{\Omega_{b}^{*}}}F_7   \,, \nonumber \\ 
		H_{\frac{3}{2}1}^{V} =& -\frac{1}{\sqrt{3}} H_{\frac{1}{2}1}^{V}(F_1,F_2,F_3) + \frac{1}{\sqrt{3}} \alpha_{\frac{1}{2}1}^{V}(\omega+1) \frac{m_{\Omega_{c}^{*}}}{m_{\Omega_{b}^{*}}}F_7   \,, \nonumber \\
		H_{\frac{1}{2}-1}^{V} =& -\frac{1}{\sqrt{3}} H_{\frac{1}{2}1}^{V}(F_1,F_2,F_3) + \frac{1}{\sqrt{3}} \alpha_{\frac{1}{2}1}^{V}(\omega+1)F_7   \,, \nonumber \\ 
		H_{\frac{3}{2}t}^{V} =& -H_{\frac{1}{2}t}^{V}(F_1,F_2,F_3)  \,, \nonumber \\
		H_{\frac{3}{2}0}^{V} =& -H_{\frac{1}{2}0}^{V}(F_1,F_2,F_3)  \,, \nonumber \\
		H_{\frac{1}{2}t}^{A} =& \frac{1-2\omega}{3} H_{\frac{1}{2}t}^{A}(F_1,F_2,F_3) + \frac{2}{3}(\omega^2-1)\frac{m_{\Omega_{c}^{*}}}{m_{\Omega_{b}^{*}}} H_{\frac{1}{2}t}^{A} (G_4,G_5,G_6)  \,, \nonumber \\
		H_{\frac{1}{2}0}^{A} =& \frac{1-2\omega}{3} H_{\frac{1}{2}0}^{A}(G_1,G_2,G_3) + \frac{2}{3}(\omega^2-1)\frac{m_{\Omega_{c}^{*}}}{m_{\Omega_{b}^{*}}} H_{\frac{1}{2}0}^{A}(G_4,G_5,G_6) + \frac{1}{3} \alpha_{\frac{1}{2}1}^{A}(\omega-1) \frac{\sqrt{2q^2}}{m_{\Omega_{b}^{*}}}G_7   \,, \nonumber \\ 
		H_{\frac{1}{2}1}^{A} =& -\frac{2\omega}{3} H_{\frac{1}{2}1}^{A}(G_1,G_2,G_3) + \frac{2}{3}(\omega^2-1)\frac{m_{\Omega_{c}^{*}}}{m_{\Omega_{b}^{*}}} H_{\frac{1}{2}1}^{A}(G_4,G_5,G_6) - \frac{1}{3} \alpha_{\frac{1}{2}1}^{A}(\omega-1) \frac{m_{-}}{m_{\Omega_{b}^{*}}}G_7   \,, \nonumber \\ 
		H_{\frac{3}{2}1}^{A} =& -\frac{1}{\sqrt{3}} H_{\frac{1}{2}1}^{A}(G_1,G_2,G_3) - \frac{1}{\sqrt{3}} \alpha_{\frac{1}{2}1}^{A}(\omega-1) \frac{m_{\Omega_{c}^{*}}}{m_{\Omega_{b}^{*}}}G_7   \,, \nonumber \\
		H_{\frac{1}{2}-1}^{A} =& \frac{1}{\sqrt{3}} H_{\frac{1}{2}1}^{A}(G_1,G_2,G_3) - \frac{1}{\sqrt{3}} \alpha_{\frac{1}{2}1}^{A}(\omega-1)G_7   \,, \nonumber \\ 
		H_{\frac{3}{2}t}^{A} =& -H_{\frac{1}{2}t}^{A}(G_1,G_2,G_3)  \,, \nonumber \\
		H_{\frac{3}{2}0}^{A} =& -H_{\frac{1}{2}0}^{A}(G_1,G_2,G_3)  \,,
\end{align}
where the following relations and definitions are employed to express the helicity amplitudes explicitly,
\begin{equation}
	\begin{split}
		H_{\frac{1}{2}t}^{V}(x,y,z) =&\ \alpha_{\frac{1}{2}t}^{V}(x\ m_{-}+z\frac{q^2}{m_{\Omega_{b}^{*}}}),\ \ \ H_{\frac{1}{2}0}^{V}(x,y,z) = \alpha_{\frac{1}{2}0}^{V}(x\ m_{+}+y\frac{q^2}{m_{\Omega_{b}^{*}}}) \,, \\
		H_{\frac{1}{2}1}^{V}(x,y,z) =& -\alpha_{\frac{1}{2}1}^{V}(x+y\frac{m_{+}}{m_{\Omega_{b}^{*}}}),\ \ \  H_{\frac{1}{2}t}^{A}(x,y,z) = \alpha_{\frac{1}{2}t}^{A}(x\ m_{+}-z\frac{q^2}{m_{\Omega_{b}^{*}}})\,, \\
		H_{\frac{1}{2}0}^{A}(x,y,z) =&\ \alpha_{\frac{1}{2}0}^{A}(x\ m_{-}-y\frac{q^2}{m_{\Omega_{b}^{*}}}),\ \ \ H_{\frac{1}{2}1}^{A}(x,y,z) = \alpha_{\frac{1}{2}1}^{A}(-x+y\frac{m_{-}}{m_{\Omega_{b}^{*}}}) \,, 
	\end{split}
\end{equation}
\begin{equation}
	\begin{split}
		m_{\pm} =&\ m_{\Omega_{b}^{*}} \pm m_{\Omega_{c}^{*}} ,\ \ \ \omega = \frac{m_{\Omega_{b}^{*}}^2+m_{\Omega_{c}^{*}}^2-q^2}{2(m_{\Omega_{b}^{*}})(m_{\Omega_{c}^{*}})} \,, \\
		\alpha_{\frac{1}{2}t}^{V} =&\ \alpha_{\frac{1}{2}0}^{A} = \sqrt{\frac{2(m_{\Omega_{b}^{*}})(m_{\Omega_{c}^{*}})(\omega+1)}{q^2}},\ \ \ \alpha_{\frac{1}{2}0}^{V} =\alpha_{\frac{1}{2}t}^{A} = \sqrt{\frac{2(m_{\Omega_{b}^{*}})(m_{\Omega_{c}^{*}})(\omega-1)}{q^2}} \,, \\  
		\alpha_{\frac{1}{2}1}^{V} =&\ 2\sqrt{(m_{\Omega_{b}^{*}})(m_{\Omega_{c}^{*}})(\omega-1)},\ \ \ \alpha_{\frac{1}{2}1}^{A} = 2\sqrt{(m_{\Omega_{b}^{*}})(m_{\Omega_{c}^{*}})(\omega+1)} \,. 
	\end{split}
\end{equation}
The helicity amplitudes corresponding to the negative helicity states for the $\frac{3}{2}^+\rightarrow \frac{3}{2}^+$ transition are also defined as,
\begin{equation}
	H_{-\lambda',-\lambda_{W}}^{V} = H_{\lambda',\lambda_{W}}^{V}\ ,\ \ \ H_{-\lambda',-\lambda_{W}}^{A} =- H_{\lambda',\lambda_{W}}^{A} \,.
\end{equation}

Finally, the decay width of the $\Omega_{b}^{*}\rightarrow\Omega_{c}^{*} \ell \bar{\nu}_{\ell}$ weak transition can be expressed as \cite{Faessler:2009xn},
\begin{equation}
	\Gamma_{\Omega_{b}^{*}\rightarrow\Omega_{c}^{*} \ell \bar{\nu}_{\ell}} = \frac{1}{2} \frac{G_{F}^{2}|V_{cb}|^2}{192\pi^3}\frac{m_{\Omega_{c}^{*}}}{m_{\Omega_{b}^{*}}^{2}} \int_{m_{\ell}^{2}}^{m_{-}^2} \frac{dq^2}{q^2} (q^2-m_{\ell}^{2})^2\ \sqrt{\omega^2-1}\ \mathcal{H}_{\frac{3}{2} \rightarrow \frac{3}{2}} \,,
\end{equation}
here $m_\ell$ denotes the mass of the leptons and $\mathcal{H}_{\frac{3}{2} \rightarrow \frac{3}{2}}$ represents the bilinear combinations of the helicity amplitudes \cite{Faessler:2009xn}:
\begin{equation}
	\label{HelAm}
	\begin{split}
	\mathcal{H}_{\frac{3}{2} \rightarrow \frac{3}{2}} =& |H_{\frac{1}{2}1}|^2 + |H_{-\frac{1}{2}-1}|^2 + |H_{\frac{3}{2}1}|^2 + |H_{-\frac{3}{2}-1}|^2 + |H_{\frac{1}{2}-1}|^2 + |H_{-\frac{1}{2}1}|^2 + |H_{\frac{1}{2}0}|^2 + |H_{-\frac{1}{2}0}|^2 \\& + |H_{\frac{3}{2}0}|^2 + |H_{-\frac{3}{2}0}|^2 + \frac{m_{\ell}^{2}}{2q^2}(3|H_{\frac{1}{2}t}|^2 + 3|H_{-\frac{1}{2}t}|^2 + 3|H_{\frac{3}{2}t}|^2 + 3|H_{-\frac{3}{2}t}|^2 + |H_{\frac{1}{2}1}|^2 + |H_{-\frac{1}{2}-1}|^2 \\& + |H_{\frac{3}{2}1}|^2 + |H_{-\frac{3}{2}-1}|^2 + |H_{\frac{1}{2}-1}|^2 + |H_{-\frac{1}{2}1}|^2 + |H_{\frac{1}{2}0}|^2 + |H_{-\frac{1}{2}0}|^2 + |H_{\frac{3}{2}0}|^2 + |H_{-\frac{3}{2}0}|^2) \,.
    \end{split}
\end{equation}
To determine the helicity amplitudes of the different components in Eq.~(\ref{HelAm}) for the weak transitions, we employ the relation $H_{\lambda',\lambda_{W}}=H_{\lambda',\lambda_{W}}^{V}-H_{\lambda',\lambda_{W}}^{A}$.

Using the above formalism, the decay widths of the different channels of the  $\Omega_{b}^{*}\rightarrow\Omega_{c}^{*} \ell \bar{\nu}_{\ell}$ weak transition are calculated numerically for the form factors corresponding to the three sets of chosen structures, and the results are summarized in Table~\ref{DeWidth}.  
\begin{table}[ht]
	\centering
	\caption{Decay widths of the $\Omega_{b}^{*}\rightarrow\Omega_{c}^{*} \ell \bar{\nu}_{\ell}$ transition for various leptonic channels, given in units of GeV}
	\begin{tabular}{cccc} 
		\toprule
		 & \textbf{$\Gamma[\Omega_{b}^{*} \rightarrow \Omega_{c}^{*} e \bar{\nu}_e]$} [GeV] & \textbf{$\Gamma[\Omega_{b}^{*} \rightarrow \Omega_{c}^{*} \mu \bar{\nu}_{\mu}]$} [GeV]  & \textbf{$\Gamma[\Omega_{b}^{*} \rightarrow \Omega_{c}^{*} \tau \bar{\nu}_{\tau}]$} [GeV] \\
		\midrule
		set 1 & $2.78_{-0.59}^{+0.75}\times10^{-12}$ & $2.77_{-0.59}^{+0.74}\times10^{-12}$ & $0.76_{-0.17}^{+0.21}\times10^{-12}$ \\
		set 2 & $2.97_{-0.53}^{+0.69}\times10^{-12}$ & $2.96_{-0.53}^{+0.68}\times10^{-12}$ & $0.86_{-0.15}^{+0.20}\times10^{-12}$ \\
		set 3 & $1.65_{-0.37}^{+0.47}\times10^{-12}$ & $1.64_{-0.37}^{+0.47}\times10^{-12}$ & $0.51_{-0.12}^{+0.15}\times10^{-12}$ \\
		average & $2.47_{-0.49}^{+0.64}\times10^{-12}$ & $2.46_{-0.49}^{+0.63}\times10^{-12}$ & $0.71_{-0.15}^{+0.19}\times10^{-12}$ \\
		\bottomrule
	\end{tabular}
	\label{DeWidth}
\end{table}
Different Lorentz structures may exhibit different sensitivities to continuum contributions, higher-dimensional condensates, and truncation effects in the operator product expansion. Therefore, considering several Lorentz-structure sets provides a useful way to estimate the systematic uncertainties associated with the structure dependence of the sum rules. In general, structures containing richer momentum dependence are expected to exhibit better numerical stability and smaller uncertainties. In the present analysis, the third Lorentz-structure set contains comparatively fewer momentum factors, which may explain its relatively smaller decay-width predictions compared to the first two sets.

In addition, we evaluated the ratio of partial decay widths between the $\tau$ channel and the $e/\mu$ channels using the data presented in Table~\ref{DeWidth},
\begin{equation}
	R = \frac{\Gamma[\Omega_{b}^{*} \rightarrow \Omega_{c}^{*} \tau \bar{\nu}_{\tau}]}{\Gamma[\Omega_{b}^{*} \rightarrow \Omega_{c}^{*} e (\mu) \bar{\nu}_{e (\mu)}]} = 0.29\pm0.01 \,.
\end{equation} 
The quoted uncertainties include the spread among the different Lorentz-structure sets considered in the present analysis. The obtained results provide QCD sum rule estimates for the semileptonic $\Omega_b^* \to \Omega_c^* \ell \bar{\nu}_{\ell}$ decay widths and the corresponding ratio of partial decay widths within the Standard Model framework. These predictions may serve as useful theoretical benchmarks for future experimental investigations of the weak decays of heavy baryons.

\section{Conclusions}
\label{con}
The study of the internal structure, properties, and decay modes of the heavy baryons has attracted increasing attention in both the theoretical and experimental investigations, particularly as their existence is progressively confirmed by the experimental data. One of the central topics in the heavy baryon physics is the study of their decay modes, particularly weak transitions, which can serve as valuable probes for uncovering potential signs of new physics. 

For the $\Omega_b^{*}$ baryon, its strong and radiative decay modes have been explored in previous works \cite{Wang:2017kfr,Ortiz-Pacheco:2023kjn,Peng:2024pyl,Aliev:2014bma,Ortiz-Pacheco:2024qcf}; however, detailed analysis of its weak decay channels remain limited. In this study, we analyzed the semileptonic weak decay of the single bottom baryon $\Omega_{b}^{*}$, specifically the transition $\Omega_{b}^{*}\rightarrow\Omega_{c}^{*} \ell \bar{\nu}_{\ell}$, across different lepton channels, employing the QCD sum rule approach. We established the sum rules for the transition form factors by computing a three-point correlation function in both the physical and QCD sides, incorporating the perturbative as well as non-perturbative contributions up to mass dimension six. In the numerical analysis, after determining the appropriate working regions of the auxiliary parameters, we obtained the $q^2$-dependent fit functions of the transition form factors within the allowed physical region of the semileptonic decay. These form factors were subsequently applied to calculate the decay widths of the weak transition for all lepton channels. 

The present pioneering investigation of the weak decays of the $\Omega_{b}^{*}$ can thus provide theoretical support and guidance for the future experimental searches, and may contribute to the identification of new decay modes. It should be emphasized that the $\Omega_{b}^{*}$ baryon, with its expected dominant radiative decay mode, presents particular experimental challenges, primarily due to the extremely soft photon in the final state. Nevertheless, we remain optimistic that forthcoming experimental programs, with enhanced sensitivity and detection techniques, will enable the study of weak transitions of the $\Omega_{b}^{*}$ baryon. Such progress would significantly enrich our understanding of the spectrum and internal dynamics of the bottom baryons, while also contributing to the definitive experimental establishment of this baryon state.
 
\section{Acknowledgments}
A.A. is thankful to L. Khajouei and M. Shekari Tousi for their useful discussions and comments. The work of P.E. and A.A. is supported by Ferdowsi University of Mashhad under the grant 4/12/63834 (1403/12/06). R. Jafariseyedabad and K. Azizi acknowledge the financial support provided by the Iran National Science Foundation (INSF) for their work under grant number 4040738. 

\appendix
\section{Final Correlation Function of the QCD side}
\label{ApenA}
The final expression of the correlation function of the theoretical side,
\begin{align}
	\label{QCDCorrF}
		&\Pi_{\rho\mu\nu}^{QCD}(p,p',q^2)= \Pi_{p_\mu p_\rho p'_\nu}^{QCD}(p^2,p'^2,q^2)\ p_\mu p_\rho p'_\nu + \Pi_{p_\rho p'_\mu p'_\nu}^{QCD}(p^2,p'^2,q^2)\ p_\rho p'_\mu p'_\nu + \Pi_{p_\rho g_{\mu\nu}}^{QCD}(p^2,p'^2,q^2)\ p_\rho g_{\mu\nu} \nonumber \\& + \Pi_{p'_\nu g_{\mu\rho}}^{QCD}(p^2,p'^2,q^2)\ p'_\nu g_{\mu\rho} + \Pi_{p_\mu g_{\nu\rho}}^{QCD}(p^2,p'^2,q^2)\ p_\mu g_{\nu\rho} +  \Pi_{p'_\mu g_{\nu\rho}}^{QCD}(p^2,p'^2,q^2)\ p'_\mu g_{\nu\rho} \nonumber \\& + \Pi_{p_\mu p_\rho p'_\nu \gamma_{5}}^{QCD}(p^2,p'^2,q^2)\ p_\mu p_\rho p'_\nu \gamma_{5} + \Pi_{p_\rho p'_\mu p'_\nu \gamma_{5}}^{QCD}(p^2,p'^2,q^2)\ p_\rho p'_\mu p'_\nu \gamma_{5} + \Pi_{p_\rho g_{\mu\nu}\gamma_{5}}^{QCD}(p^2,p'^2,q^2)\ p_\rho g_{\mu\nu}\gamma_{5} \nonumber \\& + \Pi_{p'_\nu g_{\mu\rho}\gamma_{5}}^{QCD}(p^2,p'^2,q^2)\ p'_\nu g_{\mu\rho}\gamma_{5} + \Pi_{p_\mu g_{\nu\rho}\gamma_{5}}^{QCD}(p^2,p'^2,q^2)\ p_\mu g_{\nu\rho}\gamma_{5} +  \Pi_{p'_\mu g_{\nu\rho}\gamma_{5}}^{QCD}(p^2,p'^2,q^2)\ p'_\mu g_{\nu\rho}\gamma_{5} \nonumber \\& + \Pi_{p_\mu p_\rho p'_\nu \slashed{p}}^{QCD}(p^2,p'^2,q^2)\ p_\mu p_\rho p'_\nu \slashed{p} + \Pi_{p_\rho p'_\mu p'_\nu \slashed{p}}^{QCD}(p^2,p'^2,q^2)\ p_\rho p'_\mu p'_\nu \slashed{p} + \Pi_{p_\rho g_{\mu\nu}\slashed{p}}^{QCD}(p^2,p'^2,q^2)\ p_\rho g_{\mu\nu}\slashed{p} \nonumber \\& + \Pi_{p'_\nu g_{\mu\rho}\slashed{p}}^{QCD}(p^2,p'^2,q^2)\ p'_\nu g_{\mu\rho}\slashed{p} + \Pi_{p_\mu g_{\nu\rho}\slashed{p}}^{QCD}(p^2,p'^2,q^2)\ p_\mu g_{\nu\rho}\slashed{p} +  \Pi_{p'_\mu g_{\nu\rho}\slashed{p}}^{QCD}(p^2,p'^2,q^2)\ p'_\mu g_{\nu\rho}\slashed{p} \nonumber \\& + \Pi_{p_\mu p_\rho p'_\nu \slashed{p'}}^{QCD}(p^2,p'^2,q^2)\ p_\mu p_\rho p'_\nu \slashed{p'} + \Pi_{p_\rho p'_\mu p'_\nu \slashed{p'}}^{QCD}(p^2,p'^2,q^2)\ p_\rho p'_\mu p'_\nu \slashed{p'} + \Pi_{p_\rho g_{\mu\nu} \slashed{p'}}^{QCD}(p^2,p'^2,q^2)\ p_\rho g_{\mu\nu} \slashed{p'} \nonumber \\& + \Pi_{p'_\nu g_{\mu\rho} \slashed{p'}}^{QCD}(p^2,p'^2,q^2)\ p'_\nu g_{\mu\rho} \slashed{p'} + \Pi_{p_\mu g_{\nu\rho} \slashed{p'}}^{QCD}(p^2,p'^2,q^2)\ p_\mu g_{\nu\rho} \slashed{p'} +  \Pi_{p'_\mu g_{\nu\rho} \slashed{p'}}^{QCD}(p^2,p'^2,q^2)\ p'_\mu g_{\nu\rho} \slashed{p'} \nonumber \\& + \Pi_{p_\rho p'_\nu \gamma_{\mu}}^{QCD}(p^2,p'^2,q^2)\ p_\rho p'_\nu \gamma_{\mu} + \Pi_{g_{\nu\rho} \gamma_{\mu}}^{QCD}(p^2,p'^2,q^2)\ g_{\nu\rho} \gamma_{\mu} + \Pi_{p_\mu p_\rho p'_\nu \slashed{p}\gamma_{5}}^{QCD}(p^2,p'^2,q^2)\ p_\mu p_\rho p'_\nu \slashed{p}\gamma_{5} \nonumber \\& + \Pi_{p_\rho p'_\mu p'_\nu \slashed{p}\gamma_{5}}^{QCD}(p^2,p'^2,q^2)\ p_\rho p'_\mu p'_\nu \slashed{p}\gamma_{5} + \Pi_{p_\rho g_{\mu\nu}\slashed{p}\gamma_{5}}^{QCD}(p^2,p'^2,q^2)\ p_\rho g_{\mu\nu}\slashed{p}\gamma_{5}  + \Pi_{p'_\nu g_{\mu\rho}\slashed{p}\gamma_{5}}^{QCD}(p^2,p'^2,q^2)\ p'_\nu g_{\mu\rho}\slashed{p}\gamma_{5} \nonumber \\& + \Pi_{p_\mu g_{\nu\rho}\slashed{p}\gamma_{5}}^{QCD}(p^2,p'^2,q^2)\ p_\mu g_{\nu\rho}\slashed{p}\gamma_{5} +  \Pi_{p'_\mu g_{\nu\rho}\slashed{p}\gamma_{5}}^{QCD}(p^2,p'^2,q^2)\ p'_\mu g_{\nu\rho}\slashed{p}\gamma_{5}  + \Pi_{p_\rho g_{\mu\nu}\slashed{p}\slashed{p'}}^{QCD}(p^2,p'^2,q^2)\ p_\rho g_{\mu\nu}\slashed{p}\slashed{p'} \nonumber \\& + \Pi_{p'_\nu g_{\mu\rho}\slashed{p}\slashed{p'}}^{QCD}(p^2,p'^2,q^2)\ p'_\nu g_{\mu\rho}\slashed{p}\slashed{p'} + \Pi_{p_\mu g_{\nu\rho}\slashed{p}\slashed{p'}}^{QCD}(p^2,p'^2,q^2)\ p_\mu g_{\nu\rho}\slashed{p}\slashed{p'}  +  \Pi_{p'_\mu g_{\nu\rho}\slashed{p}\slashed{p'}}^{QCD}(p^2,p'^2,q^2)\ p'_\mu g_{\nu\rho}\slashed{p}\slashed{p'} \nonumber \\& + \Pi_{p_\mu p_\rho p'_\nu \slashed{p'} \gamma_{5}}^{QCD}(p^2,p'^2,q^2)\ p_\mu p_\rho p'_\nu \slashed{p'}\gamma_{5} + \Pi_{p_\rho p'_\mu p'_\nu \slashed{p'}\gamma_{5}}^{QCD}(p^2,p'^2,q^2)\ p_\rho p'_\mu p'_\nu \slashed{p'}\gamma_{5} + \Pi_{p_\rho g_{\mu\nu} \slashed{p'}\gamma_{5}}^{QCD}(p^2,p'^2,q^2)\ p_\rho g_{\mu\nu} \slashed{p'}\gamma_{5} \nonumber \\& + \Pi_{p'_\nu g_{\mu\rho} \slashed{p'}\gamma_{5}}^{QCD}(p^2,p'^2,q^2)\ p'_\nu g_{\mu\rho} \slashed{p'}\gamma_{5} + \Pi_{p_\mu g_{\nu\rho} \slashed{p'}\gamma_{5}}^{QCD}(p^2,p'^2,q^2)\ p_\mu g_{\nu\rho} \slashed{p'}\gamma_{5} +  \Pi_{p'_\mu g_{\nu\rho} \slashed{p'}\gamma_{5}}^{QCD}(p^2,p'^2,q^2)\ p'_\mu g_{\nu\rho} \slashed{p'}\gamma_{5} \nonumber \\& + \Pi_{p_\rho p'_\nu \gamma_{\mu}\gamma_{5}}^{QCD}(p^2,p'^2,q^2)\ p_\rho p'_\nu \gamma_{\mu}\gamma_{5} + \Pi_{g_{\nu\rho} \gamma_{\mu}\gamma_{5}}^{QCD}(p^2,p'^2,q^2)\ g_{\nu\rho} \gamma_{\mu}\gamma_{5} + \Pi_{p_\rho p'_\nu \gamma_{\mu}\slashed{p}}^{QCD}(p^2,p'^2,q^2)\ p_\rho p'_\nu \gamma_{\mu}\slashed{p} \nonumber \\& + \Pi_{g_{\nu\rho} \gamma_{\mu}\slashed{p}}^{QCD}(p^2,p'^2,q^2)\ g_{\nu\rho} \gamma_{\mu} \slashed{p} + \Pi_{p_\rho p'_\nu \gamma_{\mu}\slashed{p'}}^{QCD}(p^2,p'^2,q^2)\ p_\rho p'_\nu \gamma_{\mu}\slashed{p'} + \Pi_{g_{\nu\rho} \gamma_{\mu}\slashed{p'}}^{QCD}(p^2,p'^2,q^2)\ g_{\nu\rho} \gamma_{\mu}\slashed{p'} \nonumber \\& + \Pi_{p_\rho g_{\mu\nu}\slashed{p}\slashed{p'}\gamma_{5}}^{QCD}(p^2,p'^2,q^2)\ p_\rho g_{\mu\nu}\slashed{p}\slashed{p'}\gamma_{5} + \Pi_{p'_\nu g_{\mu\rho}\slashed{p}\slashed{p'}\gamma_{5}}^{QCD}(p^2,p'^2,q^2)\ p'_\nu g_{\mu\rho}\slashed{p}\slashed{p'}\gamma_{5} + \Pi_{p_\mu g_{\nu\rho}\slashed{p}\slashed{p'}\gamma_{5}}^{QCD}(p^2,p'^2,q^2)\ p_\mu g_{\nu\rho}\slashed{p}\slashed{p'}\gamma_{5}  \nonumber \\& +  \Pi_{p'_\mu g_{\nu\rho}\slashed{p}\slashed{p'}\gamma_{5}}^{QCD}(p^2,p'^2,q^2)\ p'_\mu g_{\nu\rho}\slashed{p}\slashed{p'}\gamma_{5} +  \Pi_{p_\rho p'_\nu \gamma_{\mu}\slashed{p}\gamma_{5}}^{QCD}(p^2,p'^2,q^2)\ p_\rho p'_\nu \gamma_{\mu}\slashed{p}\gamma_{5}  + \Pi_{g_{\nu\rho} \gamma_{\mu}\slashed{p}\gamma_{5}}^{QCD}(p^2,p'^2,q^2)\ g_{\nu\rho} \gamma_{\mu} \slashed{p}\gamma_{5} \nonumber \\&  + \Pi_{p_\rho p'_\nu \gamma_{\mu}\slashed{p}\slashed{p'}}^{QCD}(p^2,p'^2,q^2)\ p_\rho p'_\nu \gamma_{\mu}\slashed{p}\slashed{p'} + \Pi_{g_{\nu\rho} \gamma_{\mu}\slashed{p}\slashed{p'}}^{QCD}(p^2,p'^2,q^2)\ g_{\nu\rho} \gamma_{\mu} \slashed{p}\slashed{p'} + \Pi_{p_\rho p'_\nu \gamma_{\mu}\slashed{p'}\gamma_{5}}^{QCD}(p^2,p'^2,q^2)\ p_\rho p'_\nu \gamma_{\mu}\slashed{p'}\gamma_{5} \nonumber \\& + \Pi_{g_{\nu\rho} \gamma_{\mu}\slashed{p'}\gamma_{5}}^{QCD}(p^2,p'^2,q^2)\ g_{\nu\rho} \gamma_{\mu}\slashed{p'}\gamma_{5} + \Pi_{p_\rho p'_\nu \gamma_{\mu}\slashed{p}\slashed{p'}\gamma_{5}}^{QCD}(p^2,p'^2,q^2)\ p_\rho p'_\nu \gamma_{\mu}\slashed{p}\slashed{p'}\gamma_{5} + \Pi_{g_{\nu\rho} \gamma_{\mu}\slashed{p}\slashed{p'}\gamma_{5}}^{QCD}(p^2,p'^2,q^2)\ g_{\nu\rho} \gamma_{\mu} \slashed{p}\slashed{p'}\gamma_{5} \,.
\end{align}

\bibliography{ref}

\bibliographystyle{utphys}

\end{document}